\documentclass[aps,pra,twocolumn,superscriptaddress,showpacs,showkeys,amsmath,amssymb]{revtex4}
\usepackage{amsfonts}
\usepackage{amssymb,amsmath}
\usepackage{mathrsfs}
\usepackage{latexsym}
\usepackage{amsmath}
\usepackage[cp1251]{inputenc}
\usepackage{graphicx}
\usepackage{dcolumn}
\usepackage{bm}
\usepackage{color}

\RequirePackage{ifthen}
\RequirePackage[pdfstartview=FitH]{hyperref}

\begin{document}

    \title{Mean-field construction for spectrum of one-dimensional Bose polaron}

    \author{G.~Panochko}
    \affiliation{College of Natural Sciences, Ivan Franko National University of Lviv,
    	107 Tarnavskyj Str., Lviv, Ukraine}
    \author{V.~Pastukhov\footnote{e-mail: volodyapastukhov@gmail.com}}
    \affiliation{Department for Theoretical Physics, Ivan Franko National University of Lviv, 12 Drahomanov Str., Lviv, Ukraine}

    \date{\today}

    \pacs{67.85.-d}

    \keywords{one-dimensional Bose polaron, path-integral approach, mean-field approximation}

    \begin{abstract}
The full momentum dependence of spectrum of a point-like impurity immersed in a dilute one-dimensional Bose gas is calculated on the mean-field level. In particular we elaborate, to the finite-momentum Bose polaron, the path-integral approach whose semi-classical approximation leads to the conventional mean-field treatment of the problem while quantum corrections can be easily accounted by standard loop expansion techniques. The extracted low-energy parameters of impurity spectrum, namely, the binding energy and the effective mass of particle, are shown to be in qualitative agreement with the results of quantum Monte Carlo simulations.
    \end{abstract}

    \maketitle

\section{Introduction}
\label{sec1}
\setcounter{equation}{0}
Recently it has been proven in \cite{Spethmann} by controlled doping of ultra-cold Rb gas with singles neutral Cs impurity atoms that Bose polarons can be formed through the interaction of a mobile impurity with the Bose-Einstein condensed ultra-cold alkali atoms. In particular, due to experimental possibility to control and change parameters of the two-body potential the impurity binding energy was measured \cite{Jorgensen} in a wide region of the boson-impurity interaction strength, the polaron life-time was evaluated \cite{Hu} in $^{87}$Rb Bose-Einstein condensate and the Rydberg polaron was observed \cite{Camargo} in a bath of $^{84}$Sr atoms.

Such an experimental progress stimulated development of the theoretical methods for studying the dynamic \cite{Grusdt_PhysRevA_97,Drescher_PhysRevA.99.023601,Lausch_PhysRevA.97.023621,Nielsen} and spectral properties of three-dimensional Bose polarons, namely, the Monte Carlo (MC) simulations \cite{Vlietinck,Pena_Ardila,Pena_Ardila2016} and various approximate theoretical approaches for calculation of parameters of the impurity spectrum \cite{Rath,Li,Novikov,Grusdt_Shchadilova,Christensen,Grusdt_Shchadilova2017,Panochko2017,Lampo}. At the same time the low-dimensional doped Bose-Einstein condensates, i.e., systems where the role of quantum fluctuations increases, are less investigated. The two-dimensional Bose polarons are not realized experimentally and a little studied \cite{Hammer,Grust2016,Grust_at_al2016,pinkster2017,Pastukhov2D} theoretically. In contrast, the situation with one-dimensional (1D) Bose polarons is much better. Remarkably that the quasi-1D trapping of small amount of impurities in Bose bath is experimentally achievable \cite{Catani_et_al} and despite of the infrared divergences in perturbative calculations \cite{Grusdt2017,Pastukhov_1D} of polaron properties, the theoretical descriptions of these systems are rapidly developing. Particularly, the mean-field (MF) considerations for uniform \cite{Astrakharchik_04,Bruderer,Volosniev,Pastukhov_2_3BI} and trapped \cite{Bruderer2012,Dehkharghani} systems, the semi-phenomenological treatments \cite{Kamenev,schecter2012,schecter2012lett,Bonart,Bonart2013} based on the Luttinger liquid approach, the essentially exact quantum \cite{Parisi} and diffusion \cite{Grusdt2017}  Monte Carlo computations in combination with the renormalisation group approach have become standard tools in the Bose polaron physics. Of course, the major peculiarity of the 1D impurity problem is the existence of exact solutions for distinct sets of the impurity atom and the surrounding Bose particles parameters \cite{McGuire_65,McGuire_66,GAMAYUN201583,Gamayun_2016,Kain_Ling_18}, and the possibility to treat the 1D Bose polaron problem analytically in the few-body \cite{Dehkharghani_mq} limit.

Most of the aforementioned theoretical efforts were directed on the investigation of low-energy parameters of the impurity spectrum and the question of the finite-momentum Bose polaron behavior is typically left opened. In general, this problem cannot be solved even on the mean-field level in higher dimensions, but the 1D case is the special one which, as it is shown below, allows for an explicit analytical solution.

\section{Formulation}
The discussed model consists of a single impurity atom immersed in the one-dimensional bath formed by $N$ Bose particles. The system is loaded in a large volume $L$ (the periodic boundary conditions are imposed) and in the following we consider only the thermodynamic limit ($N\to \infty$, $L\to \infty$ but $N/L=\bar{n}=\rm{const}$). Both the boson-boson and boson-impurity interactions are assumed to be a short-ranged (delta-like in our case) with the coupling constants $g$ and $\tilde{g}$, respectively. Adopting the second-quantized description for bosons, the appropriate Hamiltonian reads
\begin{eqnarray}\label{H}
H=-\frac{\hbar^2}{2m_I}\partial^2_{x_I}+H_B+H_{\rm int},
\end{eqnarray}
where the first term is a kinetic energy of impurity and the second one describes repulsively interacting Bose particles with mass $m$ and chemical potential $\mu$ that keeps their number fixed,
\begin{eqnarray}\label{H_0}
H_B=\int dx\, \psi^+(x)\left\{-\frac{\hbar^2}{2m}\partial^2_x
-\mu\right\}\psi(x)\nonumber\\
+\frac{g}{2}\int dx [\psi^+(x)]^2[\psi(x)]^2.
\end{eqnarray}
The third term of (\ref{H}) is the boson-impurity potential energy
\begin{eqnarray}\label{H_int}
H_{\rm int}=\tilde{g}\psi^{+}(x_I)\psi(x_I),
\end{eqnarray}
and the field operators $\psi^{+}(x)$ and $\psi(x)$ satisfy usual bosonic commutation relations $[\psi(x), \psi^+(x')]=\delta(x-x')$, $[\psi(x), \psi(x')]=0$. The further program has become a standard routine in the polaron problem: by using the continuous translational invariance of the system we perform the Lee-Low-Pines \cite{LLP} unitary transformation $H'=U^{+}HU$ with generator $U=\exp\left\{ix_Ip- x_I\int dx\, \psi^+(x)\partial_x\psi(x)\right\}$, where $\hbar p$ should be identified with the momentum of impurity. After this the Hamiltonian $H'$ commutes with the impurity momentum operator which eigenvalue can be chosen arbitrarily (zero in our case). Therefore, below we will work with the following energy operator $H'$
\begin{eqnarray}\label{H_prime}
H'=\frac{\hbar^2p^2}{2m_I}+H'_B+H'_{\rm int}+\Delta H',
\end{eqnarray}
where $H'_B=H_B|_{m\to m_r}$ is the Bose system Hamiltonian with the replacement of the bare boson mass by the impurity-boson reduced mass $m_r=mm_I/(m+m_I)$ and $H'_{\rm int}=\tilde{g}\psi^{+}(0)\psi(0)$.
The last term $\Delta H'$ represents the effective external potential and two-body interaction associated with the impurity motion
\begin{align}\label{DH}  
 &\Delta H'=i\frac{\hbar^2p}{m_I}\int dx\,\psi^{+}(x)\partial_x\psi(x)\nonumber\\
 & -\frac{\hbar^2}{2m_I}\int dx dx'\psi^{+}(x)\psi^{+}(x')\partial_{x'}\psi(x')\partial_x\psi(x).
\end{align}

In the following study we adopt the imaginary time path-integral approach. This method, as it will be shown below, is well-suited for our problem and allows to consider the properties of a system in the finite temperature region. The general procedure for passing to the path-integral formulation is perfectly described in textbooks \cite{Negele}. In order to build the Euclidean action that determines the probability amplitude one has to replace the field operators in the normal-ordered Hamiltonian by complex periodic over the imaginary time variable $\tau \in [0,\beta]$ (where $\beta$ is the inverse temperature) functions $\psi^{*}(\tau,x)$ and $\psi(\tau,x)$. Additionally we have determined the auxiliary $\tau$-dependent real variable $v(\tau)$ that splits out by means of the Hubbard-Stratonovich transformation the product in last term of (\ref{DH}) and absorbs the first one. It is also convenient to use a phase-density representation $\psi^{*}(\tau,x)=\sqrt{n(\tau,x)}e^{-i\phi(\tau,x)}$ and $\psi(\tau,x)=\sqrt{n(\tau,x)}e^{i\phi(\tau,x)}$ instead of original couple of complex fields, especially for low-dimensional systems. This substitution leads to the following action:
\begin{align}\label{S}  
& S=-\int d\tau \left(\frac{m_Iv^2}{2}+i\hbar pv\right)\nonumber\\
&+\int d\tau dx \left\{ni(\partial_{\tau}+\hbar v\partial_{x})\phi+\mu n-\tilde{g}n\delta(x) \right\}\nonumber\\
&-\frac{1}{2}\int d\tau dx \left\{\frac{\hbar^2}{m_r}n(\partial_{x}\phi)^2+\frac{\hbar^2}{4 m_r}\frac{(\partial_{x}n)^2}{n}+gn^2\right\},
\end{align}
where the dependence on $\tau$ and $x$ near appropriate fields is not explicitly written. The whole further analysis is based on the above action. Particularly it contains all important information about the low-temperature finite-momentum behavior of Bose polaron, allows the perturbative calculations in terms of boson-impurity coupling constant $\tilde{g}$ as well as the straightforward generalizations on higher dimensions. It should be noted that $S$ could be also obtained from the conventional Feynman approach applied for polaron loaded in Bose system. In this case, however, one deals with the nonlinear change of `variables' in the path-integral which leads to some discrepancy with the original result (\ref{S}). The difference is in the replacement of the reduced mass by the bare mass of bosons $m_r\to m$ in Eq.~(\ref{S}) and is fully associated with the normal ordering of operators in the second term of $\Delta H'$ (Eq.~(\ref{DH})). The MF approximation in the path-integral approach is obtained from the `classical' trajectories of the system (which are derived from least action principle $\delta S=0$), nevertheless it does not particularly mean that the MF behavior is unaffected by the quantum corrections. But the latter are accounted in the so-called Local Density Approximation (LDA) and the simplest impact of quantum fluctuations appears only on the one-loop level. 

For the MF calculations with assumption of diluteness of Bose system we can neglect the explicit imaginary time dependence in all fields $v$, $n$ and $\phi$. Additionally, by integrating out the phase fields and after some rearrangements which can be achieved only in 1D we derived the effective action
\begin{align}\label{S_MF_v}  
& \frac{S_{\rm MF}}{\beta}=-\frac{m_Iv^2}{2}-i\hbar pv+\int dx \left\{\mu-\tilde{g}\delta(x) \right\}n\nonumber\\
&-\frac{1}{2}\int dx\left\{m_rv^2\frac{(\Delta n)^2}{n}+\frac{\hbar^2}{4m_r}\frac{(\partial_{x}n)^2}{n}+gn^2\right\},
\end{align}
where $\Delta n(x)=n(x)-\bar{n}$ (with $\bar{n}=\frac{1}{L}\int dx\, n(x)$ being the average density of Bose system). At this stage the integration over $v$-variable can be performed with the result
\begin{align}\label{S_MF_p}  
&\frac{S_{\rm MF}}{\beta}=\int dx \left\{\mu n-\tilde{g}\delta(x)n-\frac{\hbar^2}{8m_r}\frac{(\partial_{x}n)^2}{n}-\frac{g}{2}n^2\right\}\nonumber\\
&-\frac{\hbar^2p^2}{2m_I}\frac{1}{1+\Delta_p},
\end{align}
where the shorthand notation $\Delta_p=\frac{m_r}{m_I}\int dx\frac{(\Delta n)^2}{n}$ is used. The forthcoming strategy is obvious: by using an equation $\delta S_{\rm MF}/\delta n=0$ we obtain the MF density profile $n_p(x)$ of Bose bath at fixed $p$ and substitute the obtained dependence in action (\ref{S_MF_p}). The resulting formula $-(S_{\rm MF}/\beta)|_{n\to n_p}$ manifests the MF ground-state contribution to grand canonical potential of $N$ bosons with the moving impurity immersed. With solution $n_p(x)$ in hands and by using the natural identity $L\bar{n}=\int dx\, n_p(x)$ we can relate the chemical potential $\mu$ to the average density of bosons and the momentum of an impurity. Then subtracting the MF energy of a pure Bose gas $L\bar{n}^2g/2$ from the total energy of system we find the polaron spectrum at zero temperature.

In general case we succeeded (see below) in the calculation of density profile $n_p(x)$ for arbitrary $p$, though, the solution for motionless impurity $n_0(x)$ is already known \cite{Volosniev} even for finite number of Bose particles. The knowledge of $n_0(x)$ allows not only to compute the Bose polaron binding energy $\varepsilon_0$ but also to evaluate the effective mass \cite{Pastukhov_2_3BI} (here we used the fact that small corrections are the same for every thermodynamic potential)
\begin{eqnarray}\label{m_star}
\frac{m^*_I}{m_I}=1+\Delta_0, \ \ \Delta_0=\frac{m_r}{m_I}\int dx \frac{[\Delta n_0(x)]^2}{n_0(x)}.
\end{eqnarray}

\begin{figure}[h!]
	\centerline{\includegraphics
		[width=0.45\textwidth,clip,angle=-0]{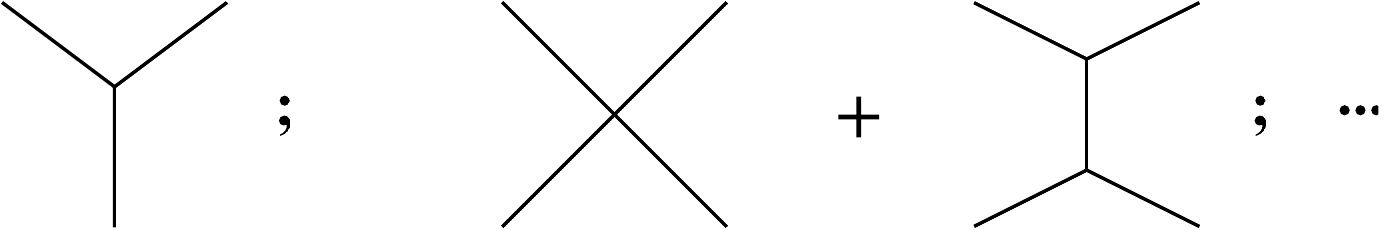}}
	\caption{Vertices determining the three- and four-particle irreducible density correlators on the `tree' level, i.e., without closed loops.}\label{Fig.1}
\end{figure}

Finally, for the limit of a weak boson-impurity interaction, where the uniform distribution of bosons is almost unperturbed by the impurity, one can use simple perturbative analysis calculating the polaron spectrum as a series in powers of $\tilde{g}$. Therefore, expanding the local density $n(x)/\bar{n}=1+\frac{1}{\sqrt{N}}\sum_{k\neq 0}e^{ikx}\rho_k$ in $S_{\rm MF}$ in terms of small deviations from the equilibrium value $\bar{n}$ we obtain the action governing the behavior of density fluctuations (from the practical point of view it is easier to work with Eq.~(\ref{S_MF_v}))
\begin{align}\label{S_MF_expanded}  
\frac{S_{\rm MF}}{\beta}=L\bar{n}\mu-\frac{1}{2}L\bar{n}^2g-\bar{n}\tilde{g} -\frac{m_Iv^2}{2}-i\hbar pv\nonumber\\
-\frac{1}{2}\sum_{k\neq 0}\left\{m_rv^2+\bar{n}g+\frac{\hbar^2k^2}{4m_r}\right\}\rho_k\rho_{-k}\nonumber\\
-\frac{1}{\sqrt{N}}\sum_{k\neq 0}\bar{n}\tilde{g}\rho_k
+\frac{1}{3!\sqrt{N}}\sum_{k+q+s= 0}\left\{3m_rv^2\right.\nonumber\\
\left.+\frac{\hbar^2}{8m_r}(k^2+q^2+s^2)\right\}\rho_k\rho_q\rho_s+\ldots,
\end{align}
where the structure of higher-order vertices can be found in Ref.~\cite{Pastukhov_15}. This perturbative consideration allows one to proceed in the canonical ensemble \cite{Pastukhov_InfraredStr}, where the chemical potential of a Bose subsystem is already fixed $\mu=\bar{n}g$ and all corrections should be associated with the impurity impact. Therefore, by integrating out $\rho_k$-fields in (\ref{S_MF_expanded}) we obtain function 
\begin{align}\label{S_v_approx}  
\frac{\mathcal{S}(v)}{\beta}= -\frac{m_Iv^2}{2}-i\hbar pv-\bar{n}\tilde{g}
+\frac{(\bar{n}\tilde{g})^2}{2N}\sum_{k\neq 0}\langle\rho_k\rho_{-k}\rangle\nonumber\\
-\frac{(\bar{n}\tilde{g})^3}{3!N^{3/2}}\sum_{k+q+s=0}\langle\rho_k\rho_q\rho_s\rangle
\pm\ldots
\end{align}
that determines polaron energy $\varepsilon_p$ at zero temperature
\begin{eqnarray}\label{e_p}
\varepsilon_p=-\lim_{\beta\to \infty}\frac{1}{\beta}\ln\left\{\int_{-\infty}^{\infty}dv\, e^{\mathcal{S}(v)}\right\},
\end{eqnarray}
perturbatively. In here notations $\langle\rho_k\rho_{-k}\rangle$, $\langle\rho_k\rho_q\rho_s\rangle$, etc. for the irreducible density correlators of pure bosons with account for the impurity flow that is represented by $v$-contributions are used. In the adopted approximation the calculations of these averages within action (\ref{S_MF_expanded}) requires the summation of a `tree' type (without internal integration over the wave-vectors) diagrams (Fig.~\ref{Fig.1}) 
for vertices with arbitrary number of external $\rho_k$-lines. When the calculations of density correlators are completed the integral in Eq.~(\ref{e_p}) should be calculated within the steepest descent method. In practice, however, one can easily evaluate only few first terms in (\ref{S_v_approx})
\begin{align}\label{S_v_as}  
&\frac{\mathcal{S}(v)}{\beta}= -\frac{m_Iv^2}{2}-i\hbar pv
-\bar{n}\tilde{g}
\left\{1-\frac{\tilde{g}}{2\hbar}\sqrt{\frac{m_r}{\bar{n}g}}\right.\nonumber\\
&\left.\times\frac{1}{\sqrt{1+m_rv^2/\bar{n}g}}
+\frac{\tilde{g}^2m_r}{6\hbar^2\bar{n}g}\frac{1+2m_rv^2/\bar{n}g}{(1+m_rv^2/\bar{n}g)^2}\pm \ldots
\right\},
\end{align}
while the result for $\varepsilon_p$ obtained within MF solution $n_p(x)$ captures summation of the whole series.

The above discussion was fully restricted to case of the Bose polaron loaded in the medium with a weak interparticle repulsion. On the mean-field level, however, parameters of the low-energy impurity spectrum depend on the LDA density profile which means that all this stuff can be easily applied to more complicated models of Bose environment. In this general case of an arbitrary boson-boson coupling strength one has to replace in Eq.~(\ref{S_MF_p}) the local MF energy density $\frac{1}{2}gn^2(x)$ by its Lieb-Liniger counterpart. Of course, the accuracy of results obtained within such an approximate calculation scheme is questionable, but it will be shown below that at least in the analytically tractable limit of infinite repulsion between Bose particles the agreement with MC simulations is quite satisfactory.

\section{Low-energy polaron spectrum}
In this section we will consider two bath-forming models, namely, the case of a dilute Bose gas and the Tonks-Girardeau (TG) limit. The characteristic feature of these examples is that in both cases the density profile $n_0(x)$ of the surrounding medium deformed by the point impurity can be written via elementary functions in the thermodynamic limit.

\subsection{Weakly-interacting Bose bath}
From the least action principle we obtain the following equation at $p=0$
\begin{eqnarray}\label{MF_densprof}
\frac{\hbar^2}{8m_r}\left(\frac{\partial_{x}n_0}{n_0}\right)^2-\frac{\hbar^2}{4m_r}\frac{\partial^2_{x}n_0}{n_0}+gn_0+\tilde{g}\delta(x)=\mu,
\end{eqnarray}
which general solution is well-discussed in literature \cite{Carr_1,Carr_2,DAgosta}. Imposing periodic boundary condition in thermodynamic limit $n_0(\pm \infty)={\rm const}$ we immediately obtain
\begin{eqnarray}  
n_0(x)=\frac{\mu}{g}\tanh^{2\sigma}(\kappa |x|+y_0),
\end{eqnarray}
here and below $\sigma={\rm sign}(\tilde{g})$, $\kappa =\sqrt{\mu m_r}/\hbar$. Parameter $y_0=\frac{1}{2}{\rm arsinh}\left(\frac{2\mu}{\kappa|\tilde{g}|}\right)$ is fixed by the boundary condition originating from the delta-function term. Now both the binding energy
\begin{eqnarray}\label{vareps_0_wc} 
\frac{\varepsilon_0}{\bar{n}g}=\frac{\bar{n}}{\kappa}\left\{\frac{4}{3}-\sqrt{1+\left(\kappa\tilde{g}/2\mu\right)^2}+\kappa\tilde{g}/2\mu\right.\nonumber\\
\left. -\frac{1}{3}\left[\sqrt{1+\left(\kappa\tilde{g}/2\mu\right)^2}-\kappa\tilde{g}/2\mu
\right]^3\right\},
\end{eqnarray}
and quantity $\Delta_0$ determining the polaron effective mass (\ref{m_star})
\begin{eqnarray}\label{Delta_0_wc}
\Delta_0=\frac{4m_r\bar{n}}{m_I\kappa}\left[\sqrt{1+\left(\kappa\tilde{g}/2\mu\right)^2}-1\right],
\end{eqnarray}
are easily calculated. In the above two formulas one can already put $\mu=\bar{n}g$. It is readily verified that the small-$\tilde{g}$ expansion of the above two equations exactly reproduce the perturbative calculations (\ref{S_MF_expanded})-(\ref{e_p}) of the impurity low-energy spectrum. On the basis of these formulas we can draw preliminary conclusions about the limits of applicability of the MF approach to the 1D Bose polaron problem. First of all, at large positive $\tilde{g}$ the binding energy asymptotically reaches constant, which is in qualitative agreement with results of MC simulations \cite{Parisi, Grusdt2017} and exact solution \cite{McGuire_65,McGuire_66} in the equal-mass limit. But for large negative values of the boson-impurity coupling parameter, $\varepsilon_0$ diverges as $-|\tilde{g}|^3$. The physics in this limit, however, is quite simple: the impurity forms a two-body bound state with particles of Bose medium and consequently the binding energy $\varepsilon_0\to -m_r\tilde{g}^2/(2\hbar^2)$. This contradiction is not actually unexpected because the structure of a MF wave function \cite{Volosniev,Pastukhov_2_3BI} does not allow for the bound-state formation in the thermodynamic limit. The calculated effective mass (\ref{Delta_0_wc}), which is found to be independent on a sign of $\tilde{g}$, reveals these shortcomings of our approach adopted to the attractive Bose polaron more clearly.

\subsection{Fermionic limit}
It is well-known that the one-dimensional Bose gas with infinite two-body $\delta$-repulsion is identical to a system of non-interacting fermions of the same density. Very often this correspondence allows to find and exact solutions of some many-body problems \cite{Yukalov}. It also simplifies our further consideration. Because in order to calculate the LDA density profile in the $g\to \infty$ limit we only have to substitute in Eq.~(\ref{MF_densprof}) the local chemical potential $\frac{\pi^2\hbar^2}{2m}n^2$ of 1D ideal Fermi gas instead of $gn$. Importantly that after this replacement the density profile of fermions in the presence of an impurity has a simple analytic form
\begin{eqnarray}  
n_0(x)=\frac{\sqrt{2m\mu}}{\pi\hbar}\frac{\cosh(2\sqrt{2}\kappa |x|+y_0)-\sigma}{\cosh(2\sqrt{2}\kappa |x|+y_0)+2\sigma},
\end{eqnarray}
in the thermodynamic limit. Notations are identical to the previously used except parameter $y_0$
\begin{align} 
&\coth^{\sigma}\left(\frac{y_0}{2}\right)=\frac{\xi}{\sqrt{3}}+\frac{2}{\sqrt{3}}\sqrt{1+\xi^2}\nonumber\\
&\times\cos\left\{\frac{1}{3}\arccos\left[\frac{\xi^3}{(1+\xi^2)^{3/2}}\right]\right\},
\end{align}
for the repulsive ($\sigma=1$) and attractive ($\sigma=-1$) Bose polarons, where additional abbreviation  $\xi=\frac{\sqrt{mm_r}\tilde{g}}{\sqrt{3}\pi\hbar^2\bar{n}}$ is introduced. The impurity binding energy is given by a simple integral with the explicitly written result
\begin{align}\label{vareps_0_TG}
&\varepsilon_0=\frac{\pi\hbar^2\bar{n}^2}{\sqrt{mm_r}} \left\{\frac{\coth^{\sigma}\left(\frac{y_0}{2}\right)}{3\coth^{2\sigma}\left(\frac{y_0}{2}\right)-1}-\frac{2\coth^{\sigma}\left(\frac{y_0}{2}\right)}{\left[3\coth^{2\sigma}\left(\frac{y_0}{2}\right)-1\right]^2}\right.\nonumber\\
&\left.+\frac{\sqrt{3}}{2}\ln\frac{\sqrt{3}+1}{\sqrt{3}-1}-\frac{\sqrt{3}}{2}\ln\frac{\sqrt{3}\coth^{\sigma}\left(\frac{y_0}{2}\right)+1}{\sqrt{3}\coth^{\sigma}\left(\frac{y_0}{2}\right)-1}\right\}.
\end{align}
Note that in the limit of infinite positive $\tilde{g}$ the impurity energy is finite, while unlike dilute Bose medium the binding energy of polaron in the TG case has a correct dependence on the boson-impurity coupling constant $\varepsilon_0\propto -\sqrt{mm_r}\tilde{g}^2/\hbar^2$ when $\tilde{g}\to -\infty$, but with wrong coefficient. The parameter responsible for the interaction-induced shift of effective mass for an impurity immersed in the TG gas reads
\begin{align}\label{Delta_0_TG}  
&\Delta_0=\frac{3\sqrt{mm_r}}{\pi m_I}\left[\coth^{\sigma}\left(\frac{y_0}{2}\right)-1\right.\nonumber\\
&\left.+\frac{1}{\sqrt{3}}\ln\frac{\sqrt{3}\coth^{\sigma}\left(\frac{y_0}{2}\right)+1}{\sqrt{3}\coth^{\sigma}\left(\frac{y_0}{2}\right)-1}-\frac{1}{\sqrt{3}}\ln\frac{\sqrt{3}+1}{\sqrt{3}-1}\right],
\end{align}
which is obviously an asymmetric function for the repulsive and attractive Bose polarons.

\subsection{Comparison with numerical methods}
\begin{figure}[h!]
	\centerline{\includegraphics
		[width=0.35\textwidth,clip,angle=-0]{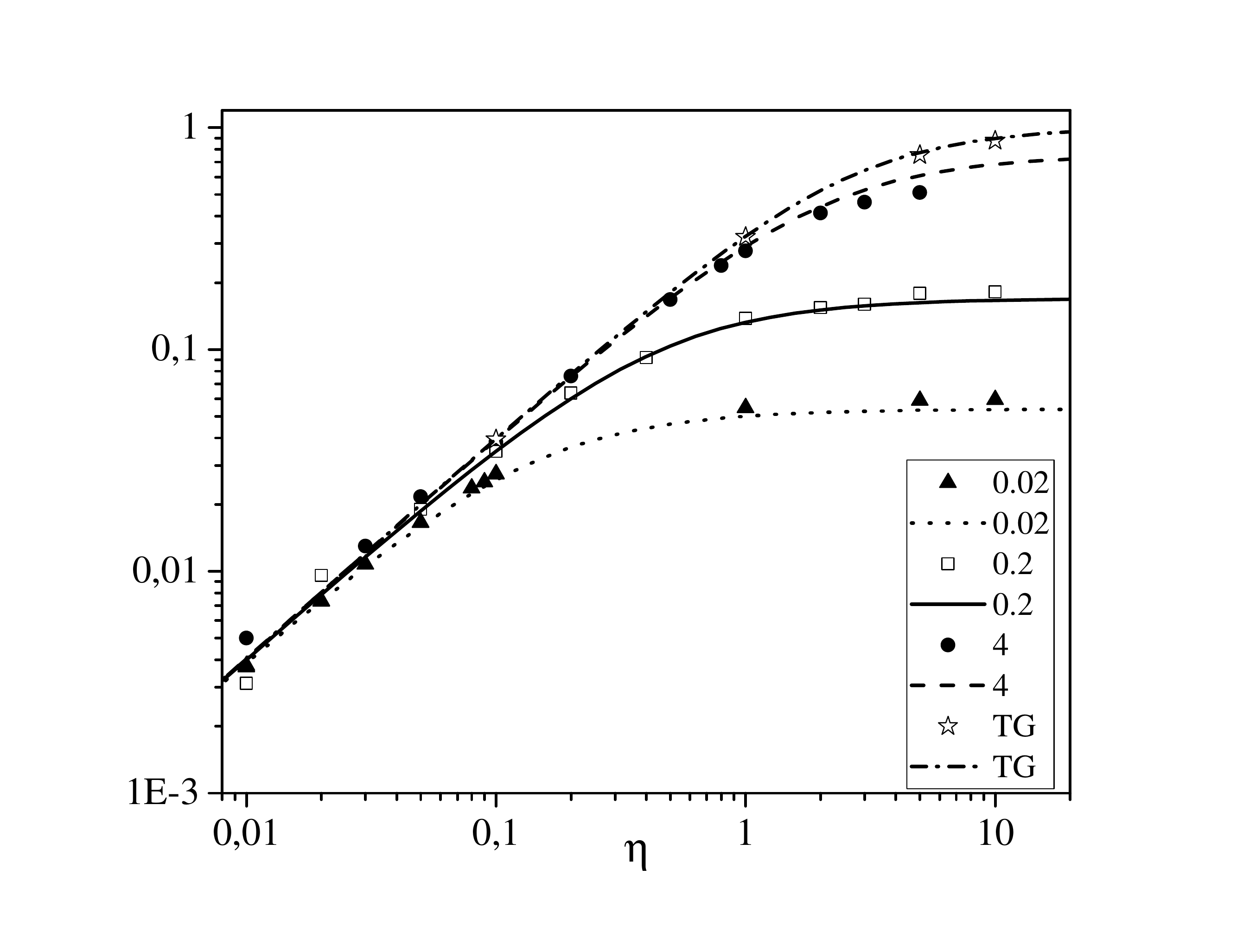}}
	\caption{Binding energy of the repulsive Bose polaron (in units of the Fermi energy) as a function of dimensionless boson-impurity coupling $\eta$ in the equal-mass limit and for $\gamma=0.02$ (dotted line), $\gamma=0.2$ (solid line), $\gamma=4$ (dashed line) and $\gamma=\infty$ (dash-dotted line). Symbols correspond to the results of quantum MC simulations \cite{Parisi}.
	\label{Fig.2}}
\end{figure}

\begin{figure}[h!]
	\centerline{\includegraphics
		[width=0.35\textwidth,clip,angle=-0]{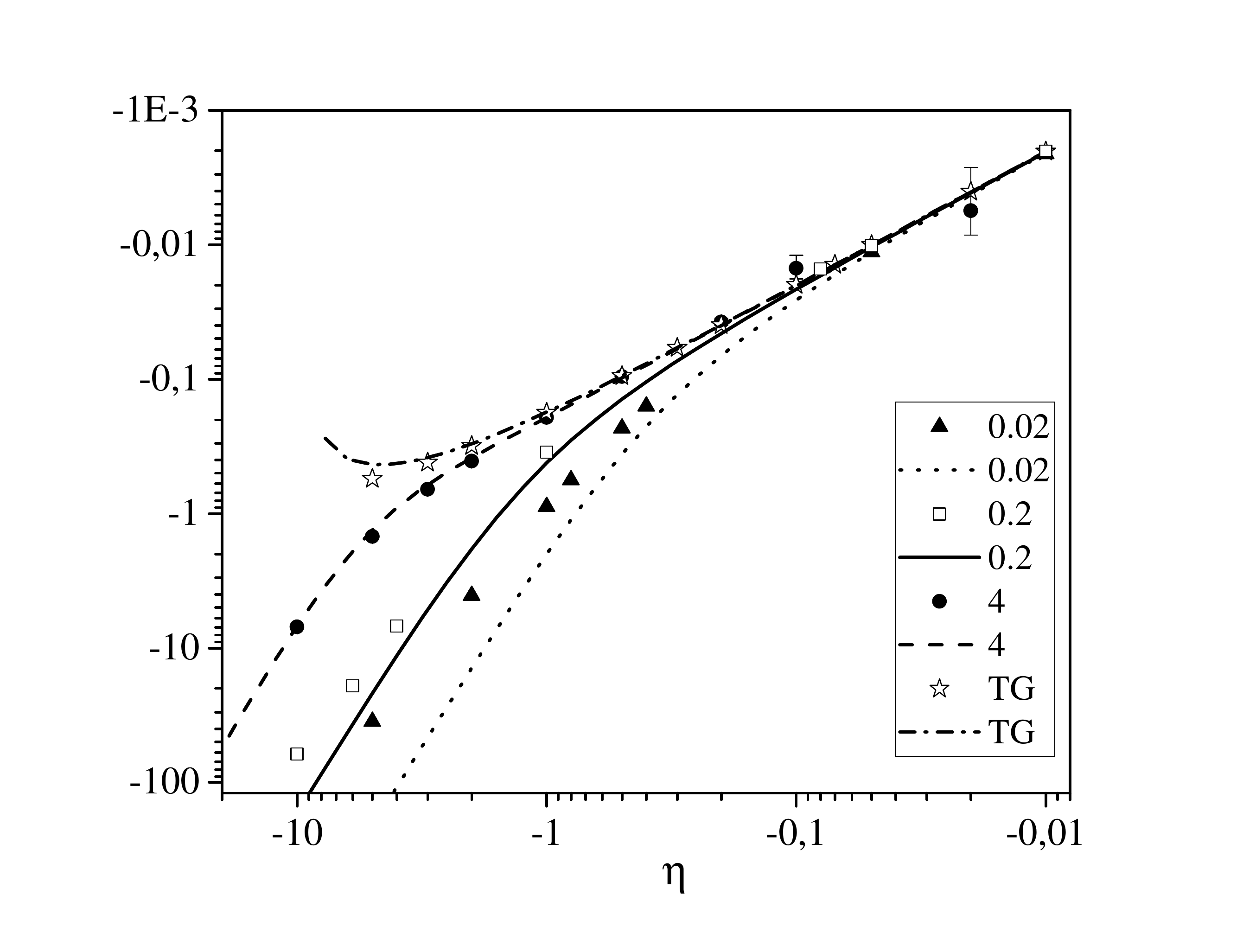}}
	\caption{Energy of the attractive mobile ($w=1$) 1D impurity shifted on the boson-impurity bound-state energy. Notations are identical to those in Fig.~\ref{Fig.2}. \label{Fig.3}}
\end{figure}

\begin{figure}[h!]	
    \includegraphics[width=0.50\textwidth,clip,angle=-0]{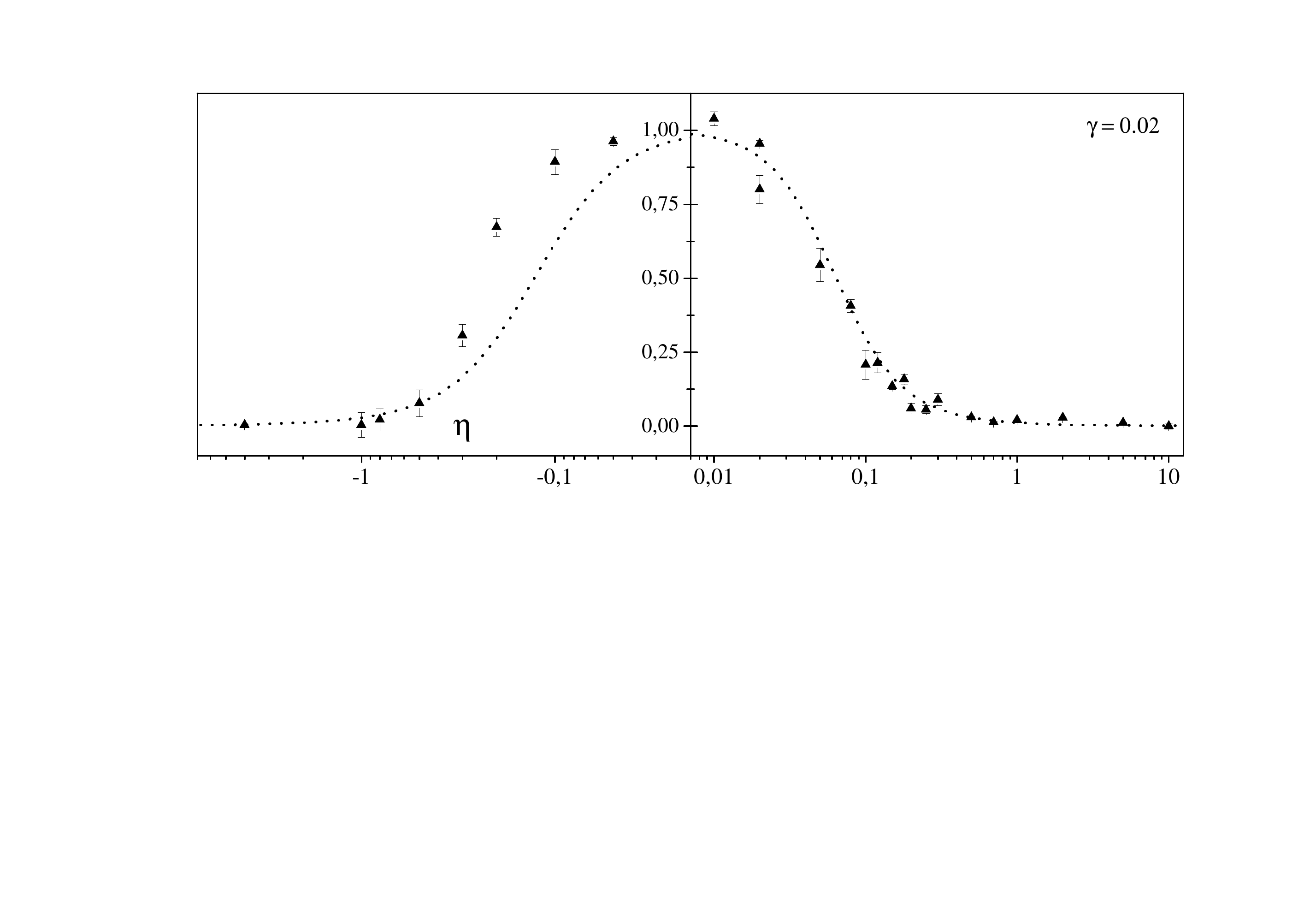}\\
    \vspace{-2.5cm}
	\includegraphics[width=0.50\textwidth,clip,angle=-0]{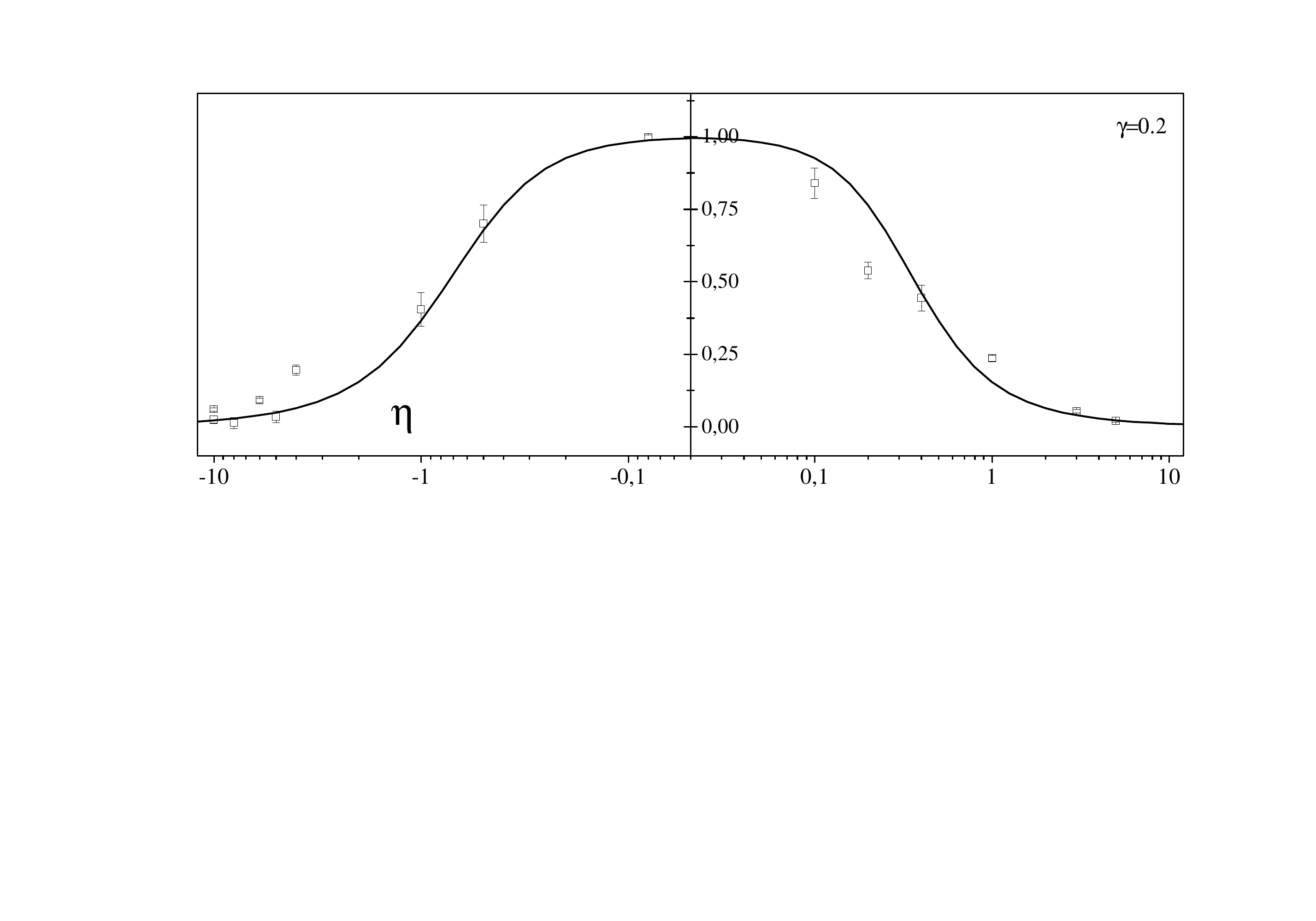}\\
	 \vspace{-2.5cm}
	\includegraphics[width=0.50\textwidth,clip,angle=-0]{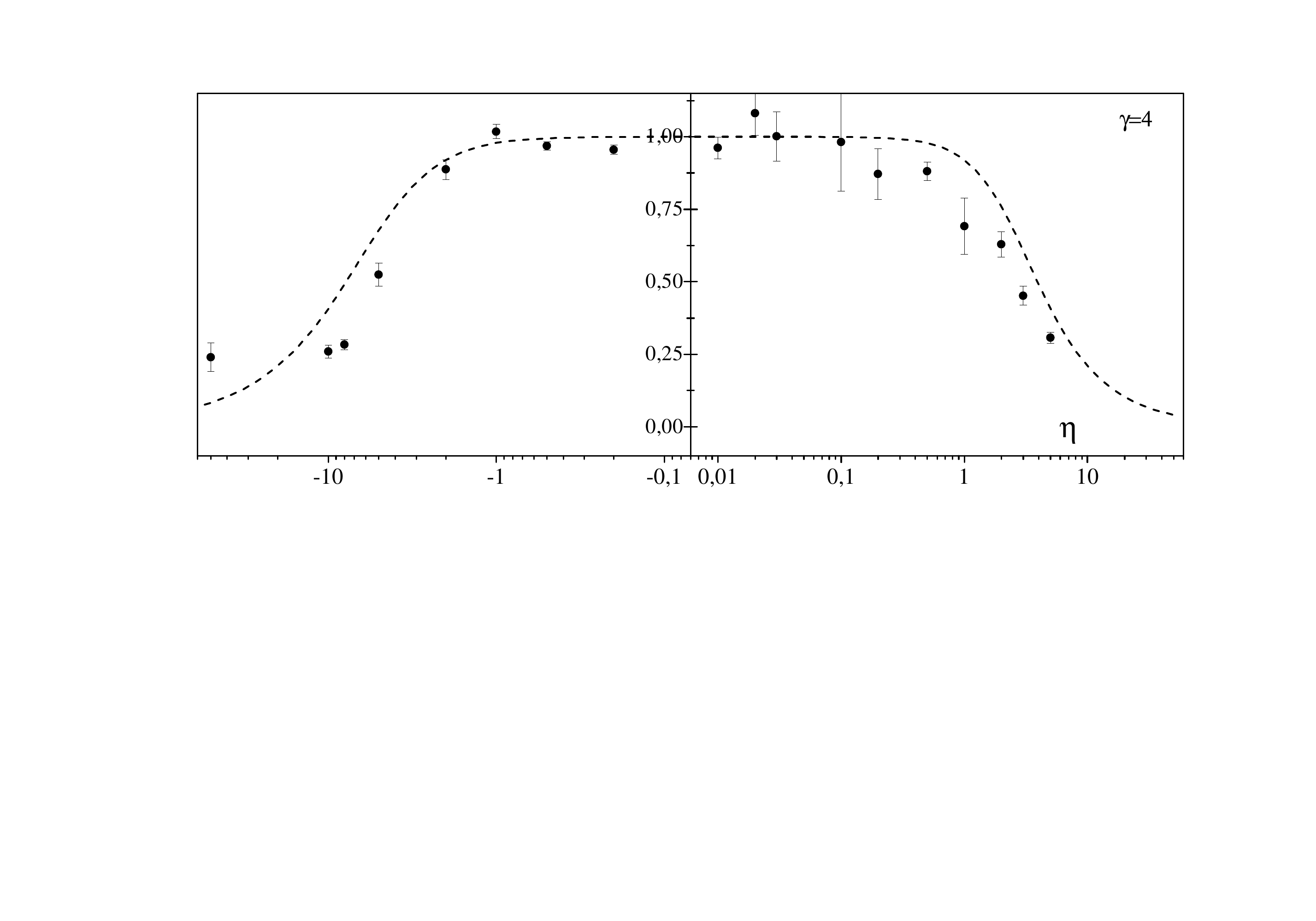}\\
	 \vspace{-2.5cm}
	\includegraphics[width=0.50\textwidth,clip,angle=-0]{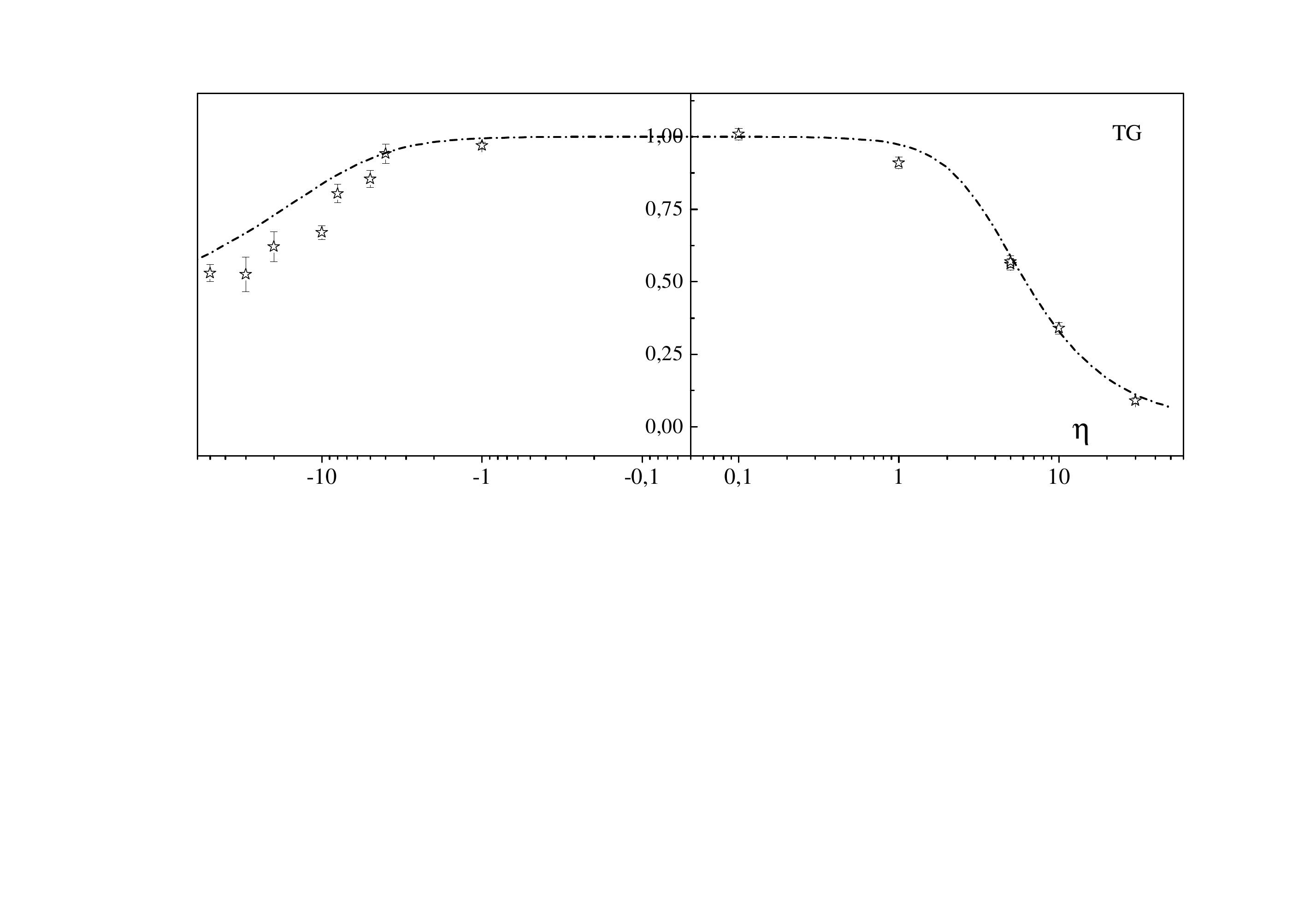}\\
	\vspace{-2.0cm}
	\caption{The inverse reduced effective mass $m_I/m^*_I$ of a mobile impurity versus $\eta$ calculated for bosonic bath in various interaction regimes: from weak (on top) to infinite (bottom) point-like repulsion. \label{Fig.4}}
\end{figure}

\begin{figure}[h!]
	\includegraphics[width=0.35\textwidth,clip,angle=-0]{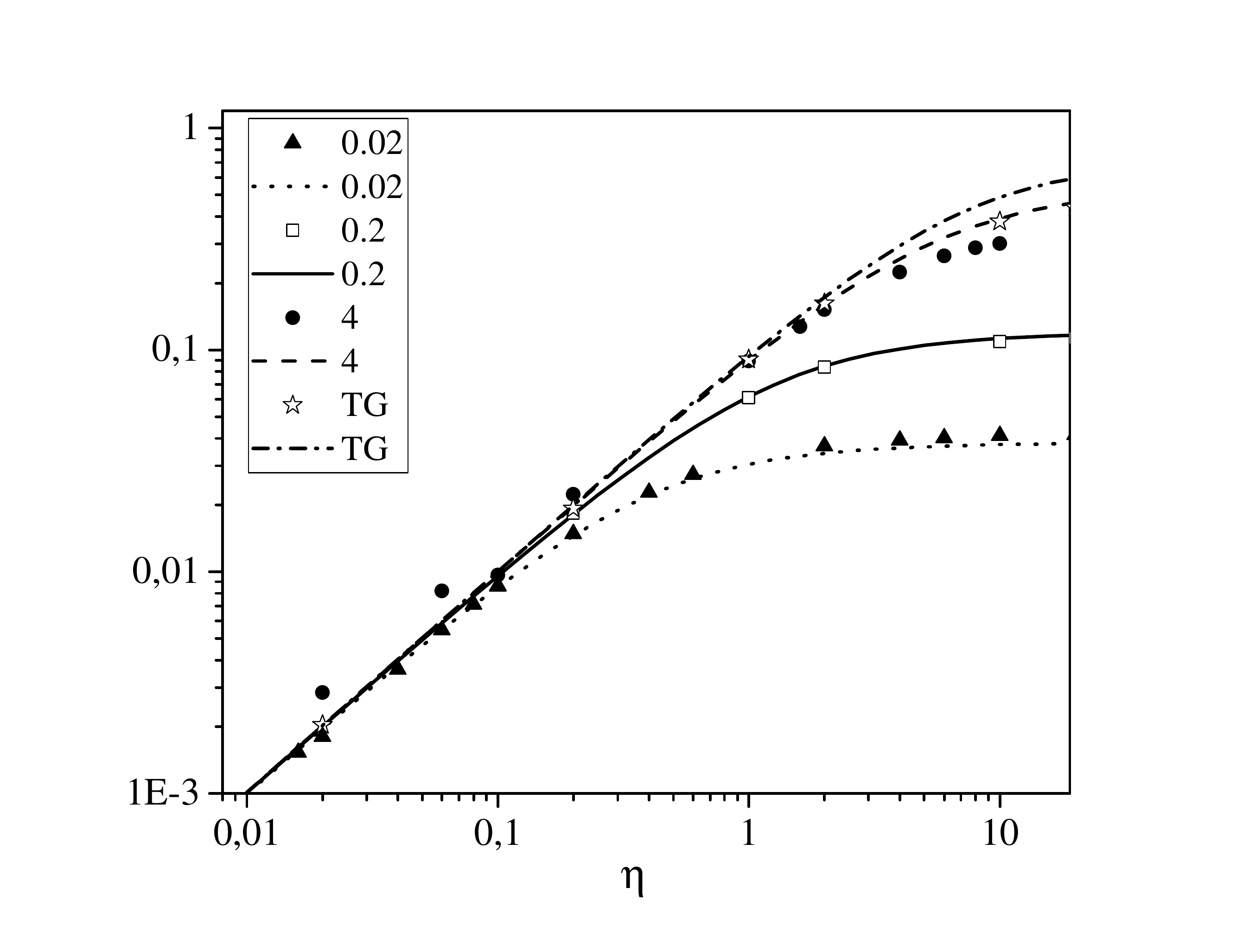}
	\includegraphics[width=0.35\textwidth,clip,angle=-0]{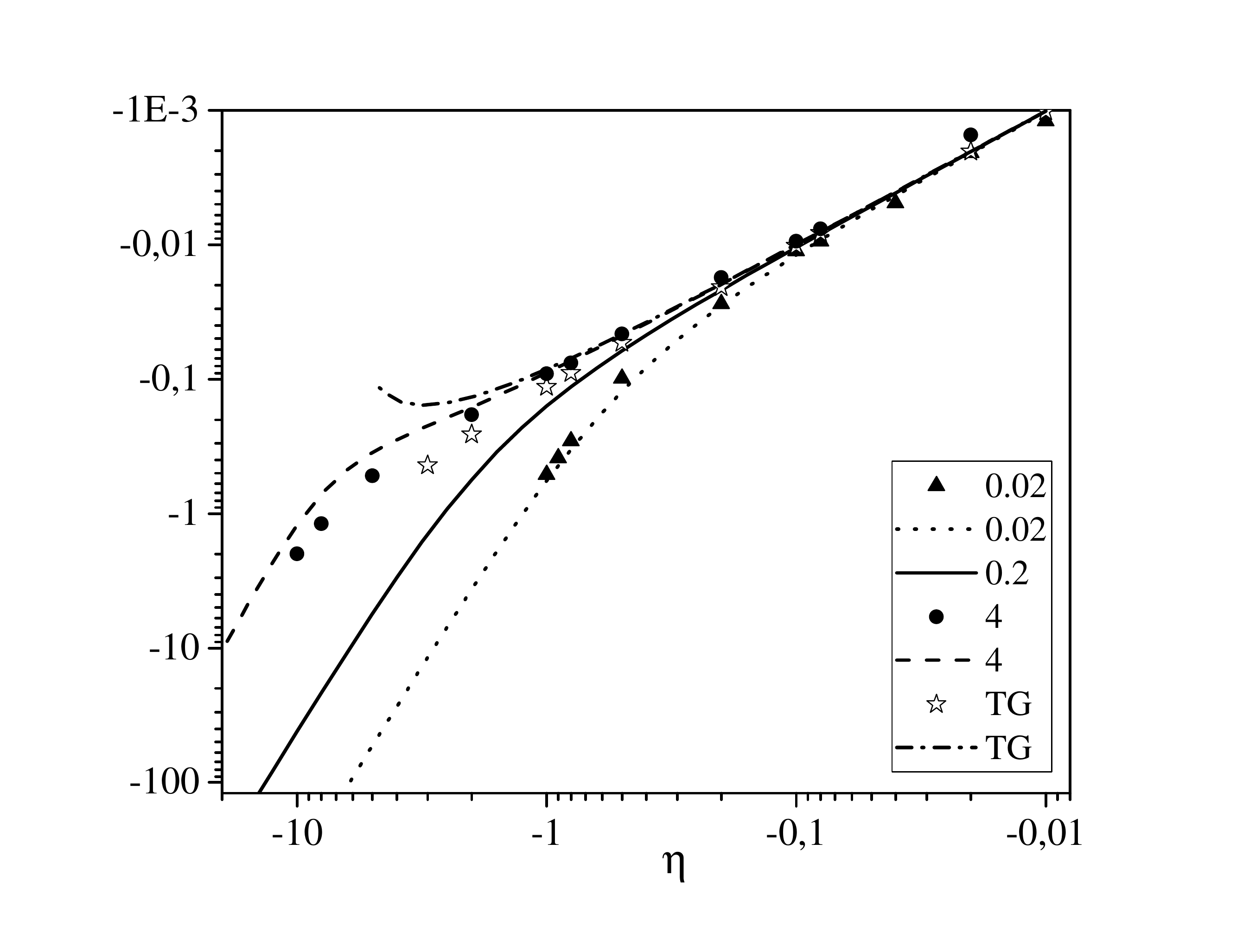}
	\caption{The dimensionless binding energy for repulsive (top) and attractive (bottom) immobile Bose polaron. Energy at negative $\eta$ is shifted by $-m\tilde{g}^2/(2\hbar^2)$ and all other designations of the plotted curves are identical to previously used. \label{Fig.5}}
\end{figure}

For comparison, we depicted in Figs.~\ref{Fig.2} the obtained curves for the low-energy parameters of polaron spectrum together with the results of quantum MC simulations \cite{Parisi}. Likewise to the above-cited paper, the dimensionless couplings are chosen as follows
\begin{eqnarray}  
	\gamma= \frac{mg}{\hbar^2\bar{n}}, \ \ \eta=\frac{2\tilde{g}m_r}{\hbar^2\bar{n}},
\end{eqnarray}
and we use notation $w=m/m_I$ for the mass ratio. First, let us consider the equal-mass limit. The binding energy (in units of the Fermi energy $\frac{\pi^2\hbar^2\bar{n}^2}{2m}$) of repulsive and attractive 1D Bose polarons versus $\eta$ for reduced couplings $\gamma=0.02$, $0.2$, $4$ and in the TG limit ($\gamma \to \infty$) are presented in Figs.~\ref{Fig.2} and Figs.~\ref{Fig.3}, respectively.
Additionally it should be stressed that the impurity energy $\varepsilon_0$ at negative $\eta$ is measured from the vacuum two-body bound-state energy $-m_r\tilde{g}^2/(2\hbar^2)$. Therefore, when the difference $\varepsilon_0+m_r\tilde{g}^2/(2\hbar^2)$ changes sign (in the TG case close to $\eta\simeq -7\div (-6)$ due to our calculations) the curve was cut off. We also compared the obtained impurity effective mass (\ref{Delta_0_wc}) (and (\ref{Delta_0_TG}) at infinite boson-boson repulsion) with the results of MC computations (see Fig.~\ref{Fig.4}). 
The same analysis was performed for the binding energy of heavy (infinite-mass) impurity. Results are depicted in Figs.~\ref{Fig.5} and we see again that the speculations of the previous subsection with the TG limit are correct only for not too intense boson-impurity attraction $\eta \ge -1$. 

Two examples considered in this section lead us to the preliminary conclusion that the MF treatment is capable to capture the qualitatively correct behavior of repulsive and weakly-coupled attractive 1D Bose polarons, while the limit of large negative $\tilde{g}$ is unattainable.

\section{Full momentum dependence}
The situation at finite impurity momenta is more interesting. Particularly, the equation that determines the Bose system density profile and provides the extremum of action (\ref{S_MF_p}) in the thermodynamic limit reads
\begin{align}\label{n_p}
&-\frac{\hbar^2}{2m_r\sqrt{n_p}}\partial^2_{x}\sqrt{n_p}+g[n_p-n^\infty_p]+\tilde{g}\delta(x)\nonumber\\
&-\frac{\hbar^2p^2}{2m_I}\frac{m_r/m_I}{(1+\Delta_p)^2}
\left[1-\left(\frac{n^{\infty}_p}{n_p}\right)^2\right]=0,
\end{align}
where abbreviation $n^{\infty}_p$ is introduced for $n_p(\pm L/2)$ (recall that $L\to \infty$), which is related to the chemical potential (up to $1/L$-terms)
\begin{eqnarray}\label{mu_p}
\mu=gn^{\infty}_p+\frac{1}{L}\frac{\hbar^2p^2}{m_I}\frac{m_r/m_I}{(1+\Delta_p)^2}
\int d x\left[1-\frac{n^{\infty}_p}{n_p}\right].
\end{eqnarray}
The latter expression together with the grand potential $-S_{\rm MF}/\beta$ enable the calculation of the MF internal energy of the system for arbitrary $p$, and consequently, the polaron spectrum
\begin{align}\label{vareps_p}
&\varepsilon_p=\frac{1}{2}g\bar{n}^2\int dx\left[1-\left(\frac{n_p}{n^{\infty}_p}\right)^2\right]
\nonumber\\
&+\frac{\hbar^2p^2/m_I}{(1+\Delta_p)^2}\left\{\frac{1}{2}+\frac{m_r}{m_I}\int d x\left[n_p-n^{\infty}_p\right]\right\}.
\end{align}
A positive solution of Eq.~(\ref{n_p}) can written in an amazingly simple analytic form \cite{Hakim}
\begin{eqnarray}  
\frac{n_p(x)}{n^{\infty}_p}=1-\frac{2\sigma(1-u^2)}{\cosh\left(2\sqrt{1-u^2}\kappa |x|+y_p\right)+\sigma},
\end{eqnarray}
where the shorthand notation $u^2=\frac{\hbar^2p^2}{m_I\mu}\frac{m_r/m_I}{(1+\Delta_p)^2}\le 1$
is adopted, while parameter $y_p$ is fixed by the boundary condition followed from the delta-term in Eq.~(\ref{n_p}):
\begin{eqnarray}\label{y_p}
\frac{\sinh(y_p)(1-u^2)^{3/2}}{[\cosh(y_p)+\sigma][\cosh(y_p)-\sigma(1-2u^2)]}=\frac{m_r|\tilde{g}|}{2\hbar^2\kappa}.
\end{eqnarray}
Now the calculation of the density profile of bosons with the immersed moving impurity is reduced to the self-consistent determination of the constants $\Delta_p$ and $y_p$ for a given set of the boson-boson and boson-impurity coupling parameters and mass ratios. And in order to find the energy $\varepsilon_p$ one has to put the obtained $n_p(x)$ under the integrals in Eq.~(\ref{vareps_p}) that can be computed to the very end with the result given by a somewhat cumbersome nevertheless explicit expression. The above equation on $y_p$ leads to the cubic one, which roots differ substantially for the attractive and repulsive Bose polarons. Therefore, these two cases should be considered separately in our further analysis. 

\begin{figure}[h!!]
	\includegraphics[width=0.35\textwidth,clip,angle=-0]{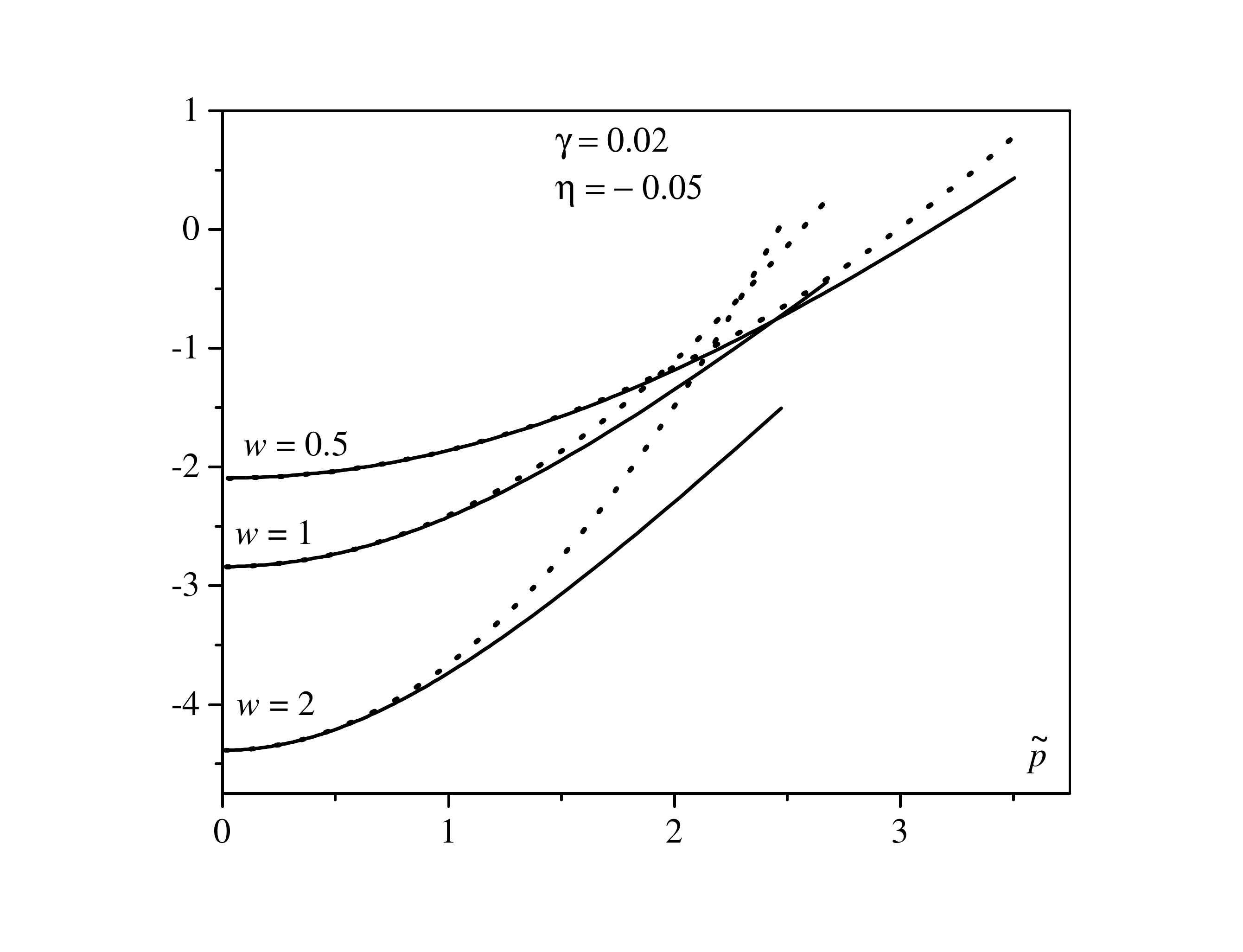}
	\includegraphics[width=0.35\textwidth,clip,angle=-0]{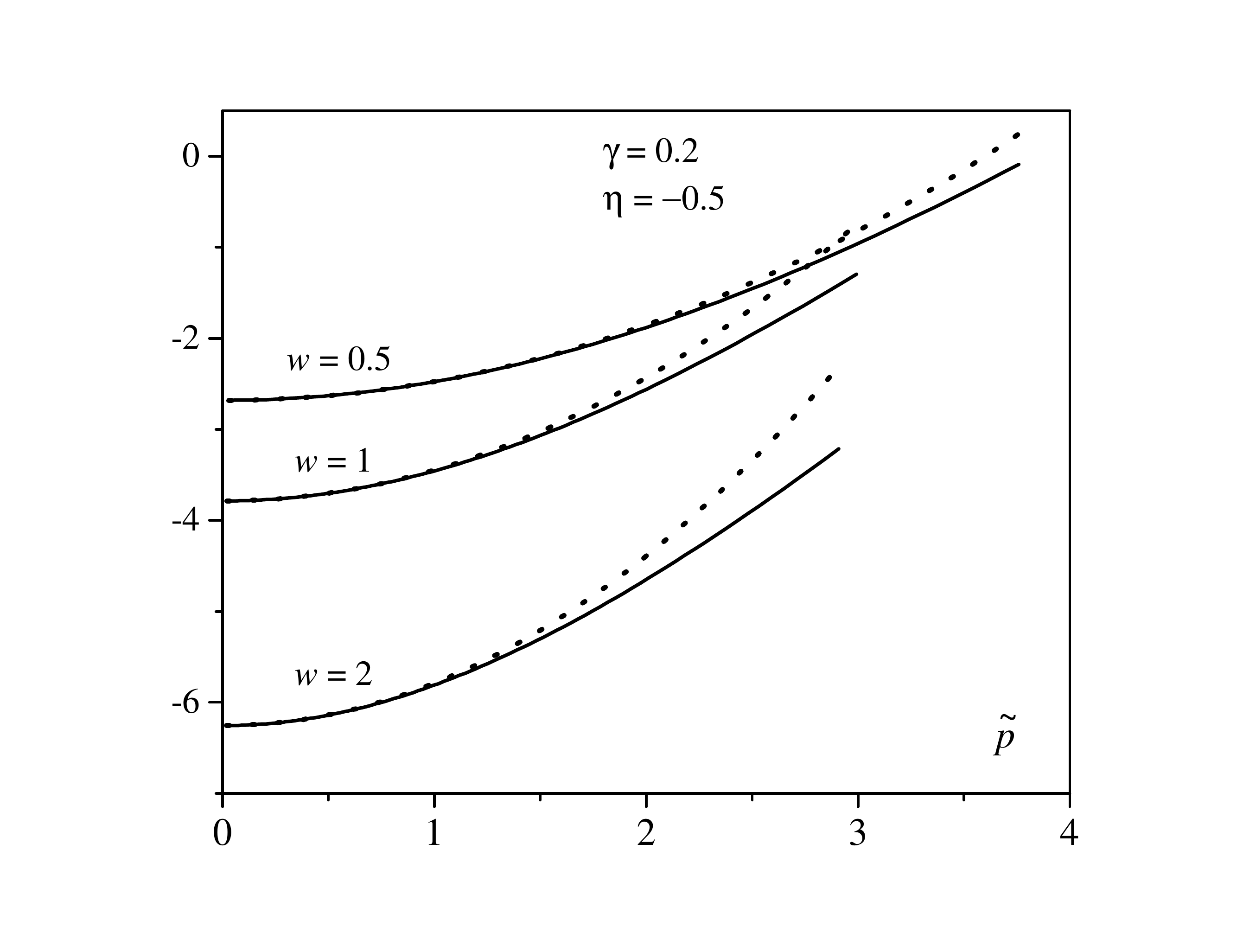}
	\includegraphics[width=0.35\textwidth,clip,angle=-0]{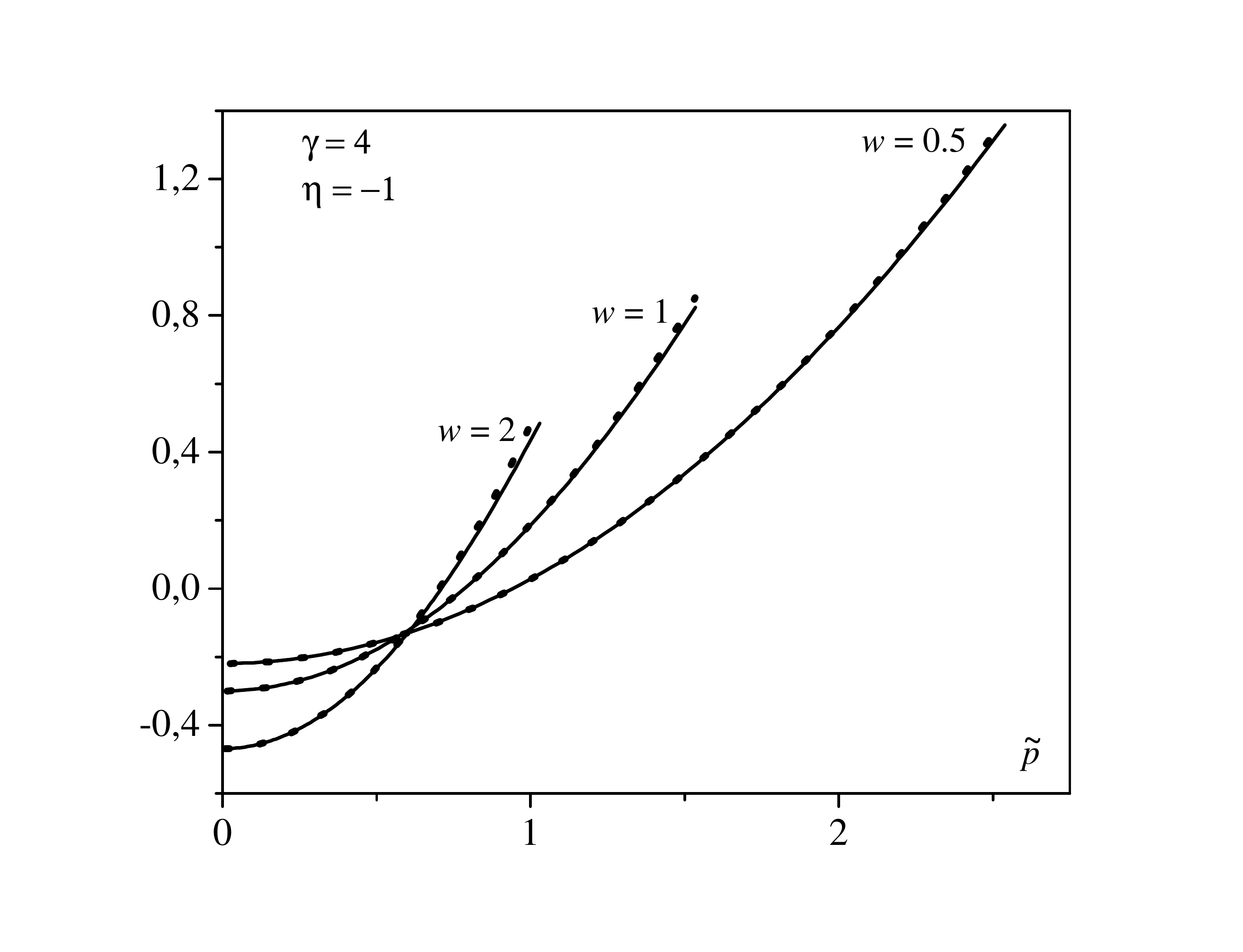}
	\caption{Momentum dependence of the dimensionless energy (in units of $\bar{n}g$) of the attractive Bose for three different mass ratios $w=1/2, 1, 2$. Dotted lines show the effective-mass approximation for the impurity dispersion relation. \label{Fig.6}}
\end{figure}

The simplest behavior is observed for negative values of parameter $\tilde{g}$. In principle, in this case the impurity that attractively interacts with the surrounding Bose particles can move through the 1D superfluid with arbitrary magnitude of velocity and the only restriction is specified by inequality $u\le 1$. Typical examples of the polaron energy in units of $\mu=\bar{n}g$ as a function of reduced momentum $\tilde{p}=\hbar p/mc$ (where $c=\sqrt{\bar{n}g/m}$ is the sound velocity in the 1D dilute Bose gas) in the attractive case is presented in Fig.~\ref{Fig.6} (solid lines).
For comparison the effective-mass approximation for the Bose polaron energy $\varepsilon_p=\varepsilon_0+\frac{\hbar^2p^2}{2m^*_I}$, where $\varepsilon_0$ is given by Eq.~(\ref{vareps_0_wc}) and $m^*_I$ determined by Eq.~(\ref{m_star}) [with $\Delta_0$ taken from (\ref{Delta_0_wc})] is also plotted (dotted lines). It is particularly seen that the parabolic behavior is a quite reasonable approximation for the impurity dispersion relation at least for attractive boson-impurity interaction.

\begin{figure}[h!!]
	\includegraphics[width=0.35\textwidth,clip,angle=-0]{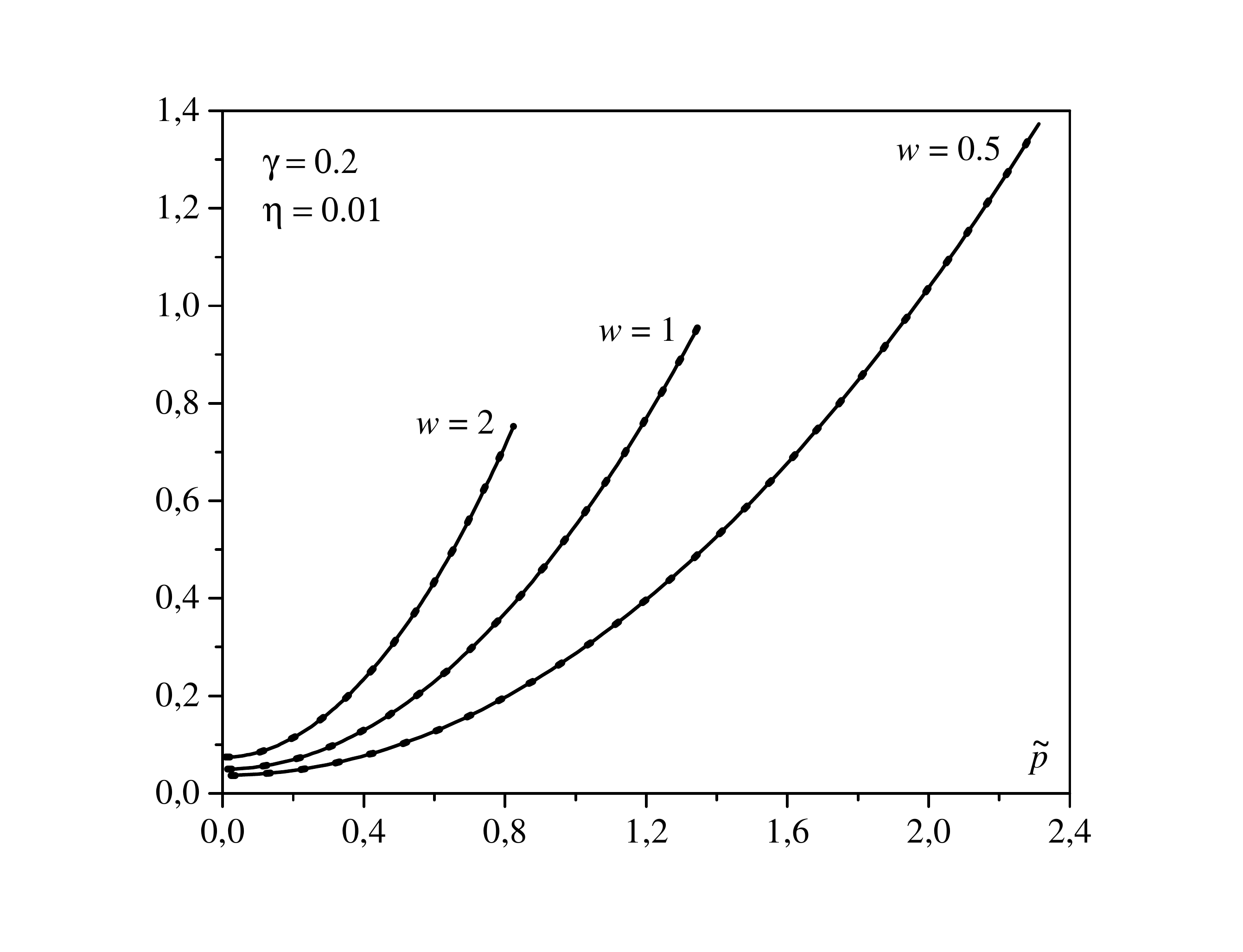}\\
	\includegraphics[width=0.35\textwidth,clip,angle=-0]{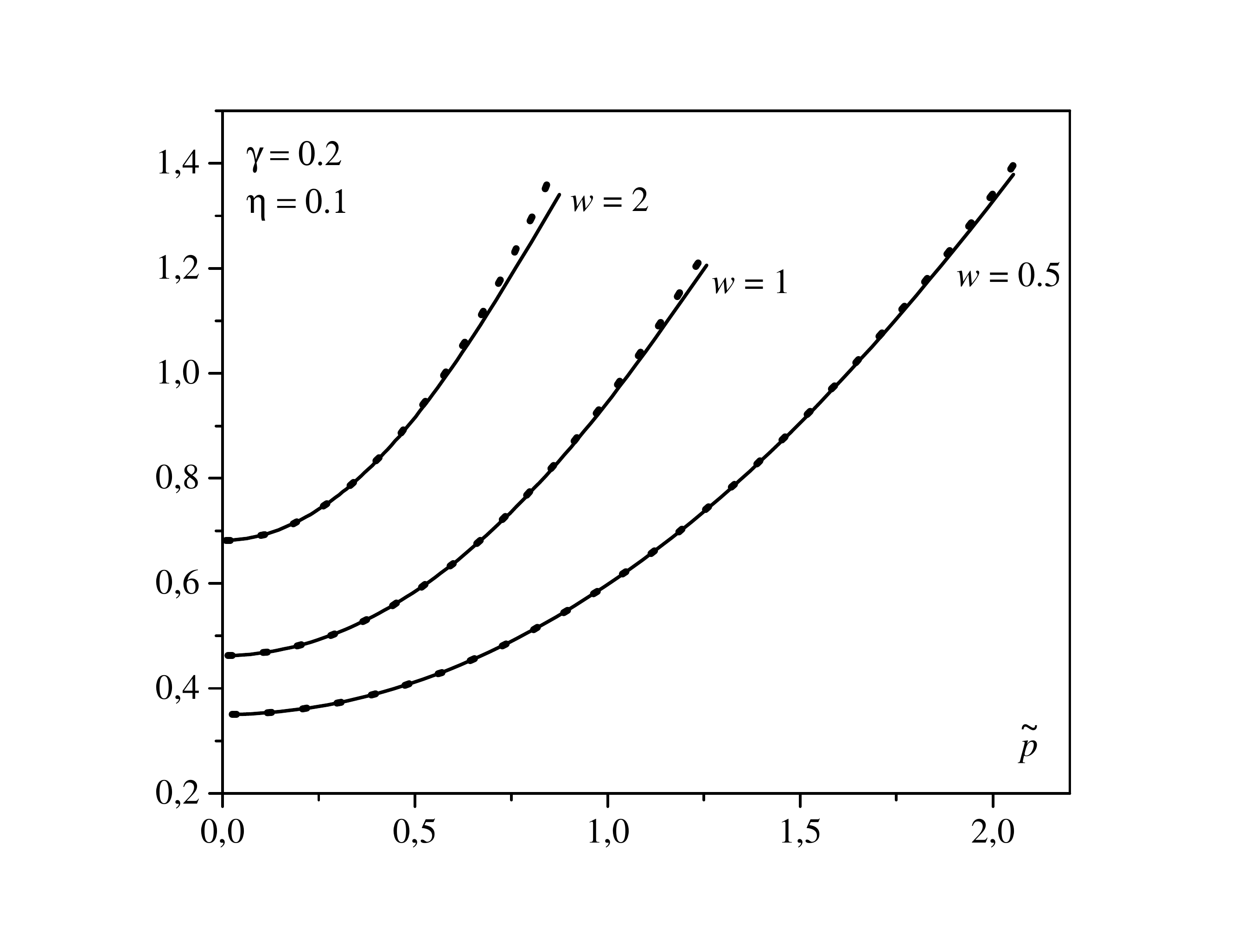}\\
	\includegraphics[width=0.35\textwidth,clip,angle=-0]{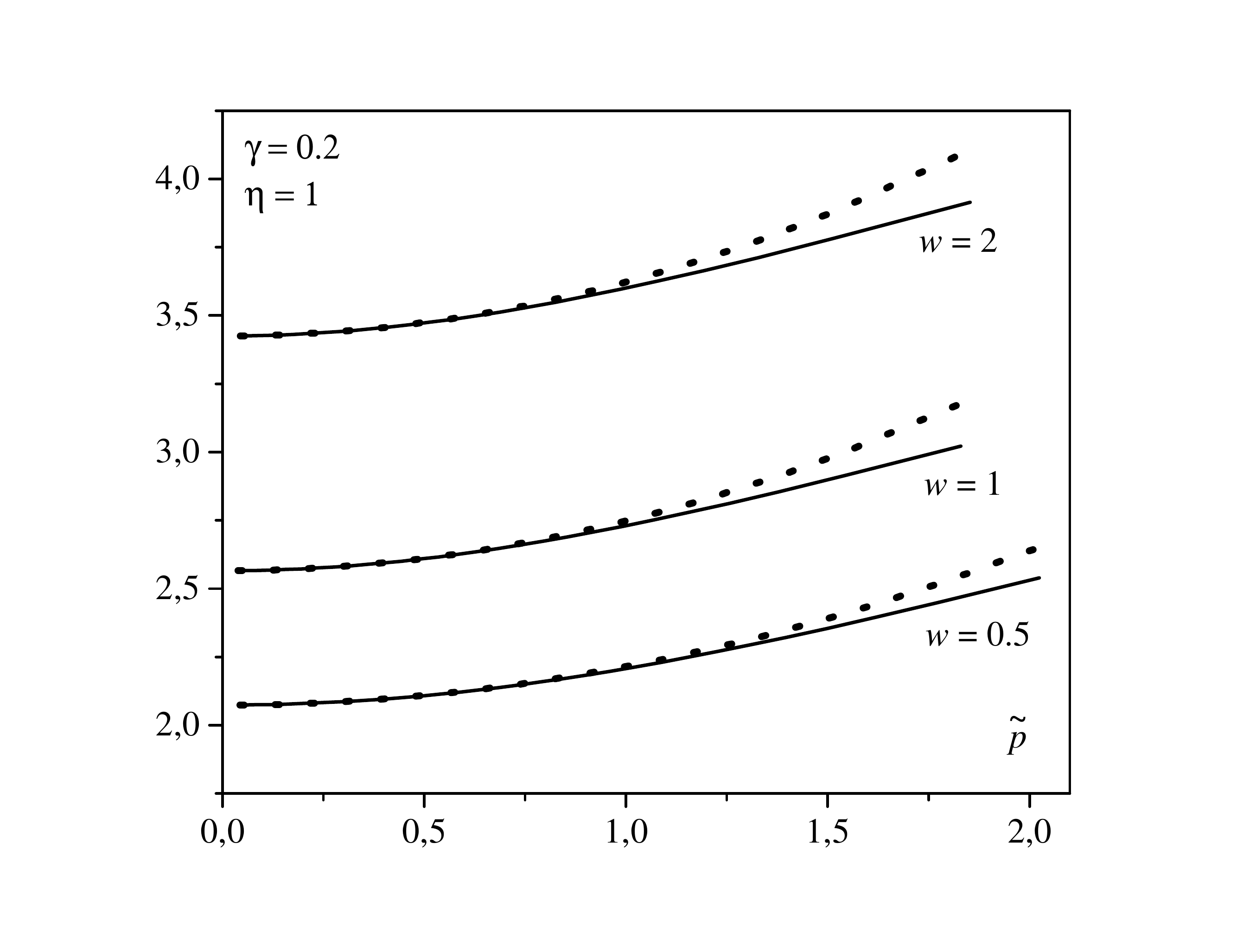}\\
	\caption{Reduced polaron energy $\varepsilon_p/(\bar{n}g)$ in the repulsive case as function of dimensionless momentum $\tilde{p}$ for three different boson-impurity interaction regimes $\eta =0.01, 0.1, 1$ (from top to bottom). Dots stand for the effective-mass approximation for an impurity spectrum. \label{Fig.7}}
\end{figure}

In the repulsive branch of the Bose polaron energy for an arbitrary set of parameters $\gamma$, $\eta$ and $w$ there always exists some `critical' value $u_c<1$ such that for $u>u_c$ Eq.~(\ref{y_p}) has no real solutions. This parameter $u_c$ also fixes the maximal value of the impurity momentum $p_c$ for which the non-uniform density $n_p(x)$ of Bose particles is a non-singular function of position. For the numerical calculations we choose the dimensionless boson-boson coupling $\gamma=0.2$ which corresponds to the dilute Bose bath (where the MF approximation is valid) and by varying the boson-impurity constant from weak $\eta=0.01$ to considerable strong $\eta=1$ repulsion we have plotted in Fig.~\ref{Fig.7} 
the momentum dependence of the Bose polaron energy for three mass ratios. Note that for any $\eta>1$ the quantitative behavior of $\varepsilon_p$ on $p$ is almost similar to that of the $\eta=1$ case. Again, the effective-mass approximation (dots) works incredibly good in all three examples. Moreover the general tendency is clearly visible: the quadratic interpolation of the original momentum-dependent polaron energy is well applicable at small $\eta$s and with the increasing of the impurity atom bare mass.

\section{Conclusions}
In summary, we have calculated, by means the of efficient path-integral method, the finite-momentum properties of a mobile point-like impurity immersed in one-dimensional Bose gas. The proposed method, which in the simplest saddle-point approximation reproduces the mean-field equation for the density profile of bath particles in the presence of moving Bose polaron and can be extended by including the beyond mean-field quantum corrections, allowed to obtain the full momentum dependence of the mean-field energy dispersion for one-dimensional Bose polaron in terms of elementary functions. The most unexpected conclusion of this study is that practically in all important cases the energy of a moving polaron can be freely approximated by the quadratic function and this observation could provide a background for developing various non-perturbative calculation techniques \cite{Panochko_18} and the  effective one-body approaches \cite{Mistakidis}. The accuracy of the presented approach is demonstrated by extracting the low-energy parameters of the impurity spectrum such as binding energy and effective mass for dilute Bose medium and in the Tonks-Girardeau case and comparing them with the results of Monte Carlo simulations. The mean-field approximation is shown to be a quite reasonable one, especially for the repulsive branch of Bose polaron energy. Finally, extension of the proposed path-integral approach may you useful in studies concerning the finite-momentum behavior of Bose polarons in higher dimensions and in the exploring of the 1D bipolaronic \cite{Dehkharghani_2018,Reichert} physics.
\begin{center}
	{\bf Acknowledgements}
\end{center}
We thank Dr.~A.~Volosniev for useful correspondence and Dr.~L.~Parisi for sharing with us the results of his Monte Carlo simulations. The authors are also grateful to Prof.~A.~Rovenchak for permanent support.


\begin{thebibliography}{99}
\bibitem{Spethmann} N.~Spethmann, F.~Kinderman, S.~John, C.~Weber, D.~Meschede, and A.~Widera, Phys.~Rev.~Lett. {\bf 109}, 235301 (2012).
\href{https://journals.aps.org/prl/abstract/10.1103/PhysRevLett.109.235301}{10.1103/PhysRevLett.109.235301}

\bibitem{Jorgensen} N.~B.~Jorgensen, L.~Wacker, K.~T.~Skalmstang, M.~M.~Parish, J.~Levinsen, R.~S.~Christensen, G.~M.~Bruun and J.~J.~Arlt,
Phys.~Rev.~Lett.  {\bf 117}, 055302 (2016).
\href{https://journals.aps.org/prl/abstract/10.1103/PhysRevLett.117.055302}{10.1103/PhysRevLett.117.055302}

\bibitem{Hu} M.-G.~Hu, M.~J.~Van de Graaff, D.~Kedar, J.~P.~Corson, E.~A.~Cornell and D.~S.~Jin, Phys.~Rev.~Lett. {\bf 117}, 055301 (2016).
\href{https://journals.aps.org/prl/abstract/10.1103/PhysRevLett.117.055301}{10.1103/PhysRevLett.117.055301}

\bibitem{Camargo} F.~Camargo, R.~Schmidt, J.~D.~Whalen, R.~Ding, G.~Woehl Jr., S.~Yoshida, J.~Burgdorfer, F.~B.~Dunning, H.~R.~Sadeghpour, E.~Demler and T.~C.~Killian, Phys.~Rev.~Lett. {\bf 120}, 088401 (2018).
\href{https://journals.aps.org/prl/abstract/10.1103/PhysRevLett.120.083401}{10.1103/PhysRevLett.120.083401}

\bibitem{Grusdt_PhysRevA_97}F.~Grusdt, K.~Seetharam,  Y.~Shchadilova, E.~Demler, Phys.~Rev.~A
 {\bf 97}, 033612 (2018).
\href{ https://journals.aps.org/pra/abstract/10.1103/PhysRevA.97.033612}{10.1103/PhysRevA.97.033612}

\bibitem{Lausch_PhysRevA.97.023621} T.~Lausch, A.~Widera, M.~ Fleischhauer Phys.~Rev.~A  {\bf 97} 023621 (2018).
\href{  https://journals.aps.org/pra/abstract/10.1103/PhysRevA.97.023621}{10.1103/PhysRevA.97.023621}

\bibitem{Drescher_PhysRevA.99.023601} M.~Drescher, M.~ Salmhofer, T.~ Enss, Phys. Rev. A {\bf 99}, 023601 (2019).
\href{https://journals.aps.org/pra/abstract/10.1103/PhysRevA.99.023601}{10.1103/PhysRevA.99.023601}


\bibitem{Nielsen} K.~ K.~ Nielsen, L.~A.~Pena Ardila, G.~ M.~Bruun, T.~ Pohl  New J. Phys. (2019).
\href{https://iopscience.iop.org/article/10.1088/1367-2630/ab0a81/meta}{10.1088/1367-2630/ab0a81}

\bibitem{Vlietinck} J.~Vlietinck, W.~Casteels, K.~V.~Houcke, J.~Tempere, J.~Ryckebusch and J.~T.~Devreese, New J.~Phys. {\bf 17}, 033023 (2015).
\href{https://iopscience.iop.org/article/10.1088/1367-2630/17/3/033023/meta}{10.1088/1367-2630/17/3/033023}

\bibitem{Pena_Ardila} L.~A.~Pena Ardila, S.~Giorgini Phys.~Rev.~A. {\bf 92}, 033612 (2015).
\href{https://journals.aps.org/pra/abstract/10.1103/PhysRevA.92.033612}{10.1103/PhysRevA.92.033612}

\bibitem{Pena_Ardila2016}  L.~A.~Pena Ardila, S.~Giorgini Phys.~ Rev.~A. {\bf 94}, 063640 (2016).
\href{https://journals.aps.org/pra/abstract/10.1103/PhysRevA.94.063640}{10.1103/PhysRevA.94.063640}

\bibitem{Ovchinnikov} A.~Novikov, M.~Ovchinnikov, J. Phys. A: Math. Theor. {\bf 42}, 135301 (2009).
\href{https://iopscience.iop.org/article/10.1088/1751-8113/42/13/135301/meta}{10.1088/1751-8113/42/13/135301}

\bibitem{Novikov} A.~Novikov, M.~Ovchinnikov,  J. Phys. B: At. Mol. Opt. Phys. {\bf 43}, 105301 (2010).
\href{https://iopscience.iop.org/article/10.1088/0953-4075/43/10/105301/meta}{10.1088/0953-4075/43/10/105301}

\bibitem{Rath} S.~P.~Rath, R.~Schmidt, Phys. Rev. A {\bf 88}, 053632 (2013).
\href{https://journals.aps.org/pra/abstract/10.1103/PhysRevA.88.053632}{10.1103/PhysRevA.88.053632}

\bibitem{Li} W.~Li, S.~Das Sarma, Phys.~Rev.~A. {\bf 90}, 013618 (2014).
\href{https://journals.aps.org/pra/abstract/10.1103/PhysRevA.90.013618}{10.1103/PhysRevA.90.013618}

\bibitem{Grusdt_Shchadilova} F.~Grusdt, Y.~E.~Shchadilova, A.~N.~Rubtsov, E.~Demler, Sci. Rep. {\bf 5}, 12124 (2015).
\href{https://www.nature.com/articles/srep12124}{10.1038/srep12124}

\bibitem{Christensen} R.~S.~Christensen, J.~Levinsen, and G.~M.~Bruun,  Phys. Rev. Lett. {\bf 115}, 160401 (2015).
\href{https://journals.aps.org/prl/abstract/10.1103/PhysRevLett.115.160401}{10.1103/PhysRevLett.115.160401}

\bibitem{Grusdt_Shchadilova2017} F.~Grusdt, R.~Schmidt, Y.~E.~Shchadilova, E.~Demler, Phys.~Rev.~A. {\bf  96},  013607 (2017).
\href{https://journals.aps.org/pra/abstract/10.1103/PhysRevA.96.013607}{10.1103/PhysRevA.96.013607}

\bibitem{Panochko2017} G.~Panochko,V.~Pastukhov, I.~Vakarchuk, Condens.~Matter~Phys. {\bf  20}, 13604 (2017).
\href{http://www.icmp.lviv.ua/journal/zbirnyk.89/13604/abstract.html}{10.5488/CMP.20.13604}

\bibitem{Lampo} A.~Lampo, S.~H.~Lim, M.~{\'{A}}.~Garc{\'{\i}}a-March, M.~Lewenstein, Quantum {\bf1}, 30 (2017).
\href{https://quantum-journal.org/papers/q-2017-09-27-30/}{10.22331/q-2017-09-27-30}

\bibitem{Hammer} A.~G.~Volosniev, H.-W.~Hammer, and N.~T.~Zinner
Phys.~Rev.~A {\bf 92}, 023623 (2015).
\href{https://journals.aps.org/pra/abstract/10.1103/PhysRevA.92.023623}{10.1103/PhysRevA.92.023623}

\bibitem{Grust2016} F.~Grusdt, Phys.~Rev.~B {\bf 93}, 144302 (2016).
\href{https://journals.aps.org/prb/abstract/10.1103/PhysRevB.93.144302}{10.1103/PhysRevB.93.144302}

\bibitem{Grust_at_al2016}  F.~Grusdt, M.~Fleischhauer, Phys.~Rev.~Lett. {\bf 116}, 053602 (2016).
\href{https://journals.aps.org/prl/abstract/10.1103/PhysRevLett.116.053602}{10.1103/PhysRevLett.116.053602}

\bibitem{pinkster2017} F.~Pinsker, Physica B: Cond.~Matt. {\bf 521}, 36 (2017).
\href{https://www.sciencedirect.com/science/article/pii/S0921452617303320}{10.1016/j.physb.2017.06.038}

\bibitem{Pastukhov2D} V.~Pastukhov, J.~Phys.~B: At.~Mol.~Opt.~Phys. {\bf 51}, 155203 (2018).
\href{https://iopscience.iop.org/article/10.1088/1361-6455/aacdcb/meta}{10.1088/1361-6455/aacdcb}



\bibitem{Catani_et_al} J.~Catani, G.~Lamporesi, D.~Naik, M.~Gring, M.~Inguscio, F.~Minardi, A.~Kantian and T.~Giamarchi, Phys.~Rev.~A {\bf 85}, 023623 (2012).
\href{https://journals.aps.org/pra/abstract/10.1103/PhysRevA.85.023623}{10.1103/PhysRevA.85.023623}

\bibitem{Grusdt2017} F.~Grusdt, G.~E.~Astrakharchik, E.~A.~Demler, New~J.~Phys. {\bf 19}, 103035 (2017).
\href{https://iopscience.iop.org/article/10.1088/1367-2630/aa8a2e/meta}{10.1088/1367-2630/aa8a2e}

\bibitem{Pastukhov_1D} V.~Pastukhov, Phys.~Rev.~A {\bf 96}, 043625 (2017).
\href{https://journals.aps.org/pra/abstract/10.1103/PhysRevA.96.043625}{10.1103/PhysRevA.96.043625}

\bibitem{Astrakharchik_04} G.~E.~Astrakharchik and L.~P.~Pitaevskii, Phys.~Rev.~A
{\bf 70}, 013608 (2004).
\href{https://journals.aps.org/pra/abstract/10.1103/PhysRevA.70.013608}{10.1103/PhysRevA.70.013608}

\bibitem{Bruderer} M. Bruderer, W. Bao and D. Jaksch, EPL {\bf 82}, 30004 (2008).
\href{https://iopscience.iop.org/article/10.1209/0295-5075/82/30004/meta}{10.1209/0295-5075/82/30004}

\bibitem{Volosniev} A.~G.~Volosniev, H.-W.~Hammer, Phys.~Rev.~A {\bf 96},
031601(R) (2017).
\href{https://journals.aps.org/pra/abstract/10.1103/PhysRevA.96.031601}{10.1103/PhysRevA.96.031601}

\bibitem{Pastukhov_2_3BI} V.~Pastukhov, arXiv preprint 
\href{https://arxiv.org/abs/1811.06281}{arXiv:1811.06281}


\bibitem{Bruderer2012}T.~H.~Johnson, M.~Bruderer, Y.~Cai, S.~R.~Clark, W.~Bao and D.~Jaksch, EPL (Europhys. Lett.) {\bf 98}, 26001 (2012).
\href{https://iopscience.iop.org/article/10.1209/0295-5075/98/26001/meta}{10.1209/0295-5075/98/26001}

\bibitem{Dehkharghani} A.~S.~Dehkharghani, A.~G.~Volosniev and N.~T.~Zinner, Phys.~Rev.~A {\bf 92}, 031601(R) (2015).
\href{https://journals.aps.org/pra/abstract/10.1103/PhysRevA.92.031601}{10.1103/PhysRevA.92.031601}

\bibitem{schecter2012} M.~Schecter, D.~M.~Gangardt, A.~Kamenev, Ann.~Phys. {\bf 327}, 639 (2012). 
\href{https://www.sciencedirect.com/science/article/pii/S0003491611001618}{10.1016/j.aop.2011.10.001}

\bibitem{schecter2012lett} M.~Schecter, A.~Kamenev, D.~M.~Gangardt and A.~Lamacraft, Phys. Rev. Lett. {\bf 108}, 207001 (2012).
\href{https://journals.aps.org/prl/abstract/10.1103/PhysRevLett.108.207001}{10.1103/PhysRevLett.108.207001}

\bibitem{Bonart} J.~Bonart and L.~F.~Cugliandolo, Phys.~Rev.~A {\bf 86}, 023636 (2012).
\href{https://journals.aps.org/pra/abstract/10.1103/PhysRevA.86.023636}{10.1103/PhysRevA.86.023636}

\bibitem{Bonart2013} J.~Bonart and L.~F.~Cugliandolo, EPL (Europhys. Lett.) {\bf 101} 16003 (2013).
\href{https://iopscience.iop.org/article/10.1209/0295-5075/101/16003/meta}{10.1209/0295-5075/101/16003}

\bibitem{Kamenev} A.~Kamenev  and L.~I.~Glazman, Phys.~Rev.~A {\bf 80}, 011603 (2009).
\href{https://journals.aps.org/pra/abstract/10.1103/PhysRevA.80.011603}{10.1103/PhysRevA.80.011603}

\bibitem{Parisi} L.~Parisi and S.~Giorgini, Phys.~Rev.~A {\bf 95}, 023619 (2017).
\href{https://journals.aps.org/pra/abstract/10.1103/PhysRevA.95.023619}{10.1103/PhysRevA.95.023619}
%\bibitem{Artem_Volosniev} A.~G.~Volosniev, Few-Body Syst. (2017).

\bibitem{McGuire_65} J.~B.~McGuire, J.~Math.~Phys. {\bf 6}, 432 (1965).
\href{https://aip.scitation.org/doi/abs/10.1063/1.1704291}{10.1063/1.1704291}

\bibitem{McGuire_66} J.~B.~McGuire, J.~Math.~Phys. {\bf 7}, 123 (1966).
\href{https://aip.scitation.org/doi/abs/10.1063/1.1704798}{10.1063/1.1704798}

\bibitem{GAMAYUN201583} O.~Gamayun, A.~G.~Pronko, M.~B.~Zvonarev, Nucl. Phys. B {\bf 892}, 83 (2015).
\href{https://www.sciencedirect.com/science/article/pii/S0550321315000073}{10.1016/j.nuclphysb.2015.01.004}

\bibitem{Gamayun_2016} O.~Gamayun, A.~G.~Pronko and M.~B.~Zvonarev, New J. Phys. {\bf 18}, 045005 (2016).
\href{https://iopscience.iop.org/article/10.1088/1367-2630/18/4/045005/meta}{10.1088/1367-2630/18/4/045005}

\bibitem{Kain_Ling_18} B.~Kain and H.~Y.~Ling, Phys.~Rev.~A {\bf 98}, 033610 (2018).
\href{https://journals.aps.org/pra/abstract/10.1103/PhysRevA.98.033610}{10.1103/PhysRevA.98.033610}

\bibitem{Dehkharghani_mq} A.~Dehkharghani, A.~Volosniev, J.~Lindgren, J.~Rotureau,
Ch.~Forssen, D.~Fedorov, A.~Jensen and N.~Zinner, Scientific Reports {\bf 5}, 10675 (2015).
\href{https://www.nature.com/articles/srep10675}{10.1038/srep10675}

\bibitem{LLP} T.~D.~Lee, F.~E.~Low, and D.~Pines, Phys. Rev. {\bf 90}, 297 (1953).
\href{https://journals.aps.org/pr/abstract/10.1103/PhysRev.90.297}{10.1103/PhysRev.90.297}

\bibitem{Negele} J. Negele, H. Orland, {\it Quantum Many-Particle Systems}, (Addison Wesley, 1988).

\bibitem{Pastukhov_15} V. Pastukhov, J. Phys. A: Math. Theor. {\bf 48}, 405002 (2015).
\href{https://iopscience.iop.org/article/10.1088/1751-8113/48/40/405002/meta}{10.1088/1751-8113/48/40/405002}

\bibitem{Pastukhov_InfraredStr} V.~Pastukhov, J.~Low Temp.~Phys. {\bf 186}, 148 (2017).
\href{https://link.springer.com/article/10.1007/s10909-016-1659-9}{10.1007/s10909-016-1659-9}

\bibitem{Carr_1} L.~D.~Carr, C.~W.~Clark and W.~P.~Reinhardt, Phys.~Rev.~A {\bf 62},
063610 (2000).
\href{https://journals.aps.org/pra/abstract/10.1103/PhysRevA.62.063610}{10.1103/PhysRevA.62.063610}

\bibitem{Carr_2} L.~D.~Carr, C.~W.~Clark and W.~P.~Reinhardt, Phys.~Rev.~A {\bf 62}, 063611 (2000).
\href{https://journals.aps.org/pra/abstract/10.1103/PhysRevA.62.063611}{10.1103/PhysRevA.62.063611}

\bibitem{DAgosta} R.~D'Agosta, B.~A.~Malomed and C.~Presilla, Phys.~Lett.~A {\bf 275}, 424 (2000).
\href{https://www.sciencedirect.com/science/article/abs/pii/S0375960100006198}{10.1016/S0375-9601(00)00619-8}

\bibitem{Yukalov} V.~I.~Yukalov and M.~D.~Girardeau, Laser Phys. Lett. {\bf 2} 375 (2005).
\href{https://iopscience.iop.org/article/10.1002/lapl.200510011/pdf}{10.1002/lapl.200510011}

\bibitem{Hakim} V.~Hakim, Phys.~Rev.~E {\bf 55}, 2835 (1997).
\href{https://journals.aps.org/pre/abstract/10.1103/PhysRevE.55.2835}{10.1103/PhysRevE.55.2835}

\bibitem{Panochko_18} G.~Panochko, V.~Pastukhov, I.~Vakarchuk, Int. J. Mod. Phys. B {\bf 32}, 1850053 (2018).
\href{https://www.worldscientific.com/doi/abs/10.1142/S0217979218500534}{10.1142/S0217979218500534}

\bibitem{Mistakidis} S. I. Mistakidis, A. G. Volosniev, N. T. Zinner, P. Schmelcher, arXiv preprint.
\href{https://arxiv.org/abs/1809.01889}{arXiv:1809.01889}

\bibitem{Dehkharghani_2018} A.~S. Dehkharghani, A.~G. Volosniev, and N.~T. Zinner, Phys. Rev. Lett. {\bf 121}, 080405 (2018).
\href{https://journals.aps.org/prl/abstract/10.1103/PhysRevLett.121.080405}{10.1103/PhysRevLett.121.080405}

\bibitem{Reichert} B.~Reichert, Z.~Ristivojevic, A.~Petkovic, arXiv preprint
\href{https://arxiv.org/abs/1806.03658}{ arXiv:1806.03658}





\end{thebibliography}
\end{document}